%% file: article.tex
\documentclass[journal,final,table,10pt]{IEEEtran}

\usepackage[utf8]{inputenc}

\usepackage{dirtytalk}
\usepackage{url}
\usepackage{tikz}
\usetikzlibrary{backgrounds, calc, arrows, automata, positioning, patterns, fit}
\usepackage{gnuplottex}
\usepackage{pgfplots}
\usepgfplotslibrary{groupplots, colormaps, patchplots}
\pgfplotsset{compat=newest}
\pgfplotsset{%
  colormap={whitegreen}{color(0cm)=(white);
  color(1cm)=(green!75!orange)}
}

\usepackage{multirow}
\usepackage{xcolor}
\usepackage{array}
\usepackage{etoolbox}
\usepackage{fixltx2e}
\usepackage[colorlinks=true, linkcolor=blue, urlcolor=blue, citecolor=red, anchorcolor=cyan]{hyperref}

\usepackage{mdframed}
\usepackage{subcaption, graphicx}
\usepackage{newtxtext,newtxmath}

\AtBeginEnvironment{table}{\sffamily}
\AtBeginEnvironment{tabular}{\sffamily}

\newcommand{\kwrd}[1]{\textsf{#1}}

\begin{document}

  \title{Overview and Challenges of Ambient Systems,\\Towards a Constructivist Approach to their Modelling}
  
  \author{Gérald~Rocher,
  Jean-Yves~Tigli,
  Stéphane~Lavirotte,
  and~Nhan~Le~Thanh%
  \thanks{G. Rocher, JY. Tigli, S. Lavirotte and N. Le Thanh where with the University Côte d'Azur and CNRS, I3S laboratory, Sophia-Antipolis, France}}


  \maketitle

  \begin{abstract}

\noindent From a closed and controlled environment, neglecting all the external disturbances, information processing systems are now exposed to the complexity and the aleas of the physical environment, open and uncontrolled. Indeed, as envisioned by Mark Weiser as early as 1991, the progresses made on wireless communications, energy storage and the miniaturization of computer components, made it possible the fusion of the physical and digital worlds. This fusion is embodied in a set of concepts such as Internet of Things, Pervasive Computing, Ubiquitous Computing, etc. From a synthesis of these different concepts, we show that beyond the simple collection of environmental data from sensors, the purpose of the information processing systems underlying these concepts is to carry out relevant actions that the processing of these data suggests in our environment. However, due to the complexity of these systems and the inability to predict the effects of their actions, the responsibility for these actions still often remains with users. Mark Weiser's vision of disappearing computing is still far from being a reality. This situation calls for an epistemological rupture that is proposed to be concretized through the systemic approach which finds its foundations in constructivism. It is no longer a question of predicting but of evaluating in vivo the effectiveness of these systems. The perspectives for such an approach are discussed.
\end{abstract}
  


\section{Introduction}\label{sec:introduction}

\noindent At the end of the 1980s, Mark Weiser laid the foundations of \textit{Ubiquitous Computing}, which he theorized in 1991 in his article \say{The Computer for the 21st century} \cite{weiser1991computer}. This paradigm consists in embedding computer resources on everyday objects and, even before the advent of the Internet, already suggests a form of omnipresent network. In 1995, with John Seely Brown, he pushed the concept to its limits by introducing the notion of \textit{Calm technology} \cite{weiser1996designing} whose founding idea can already be found in his 1991 seminal article: \say{\textit{The most profound technologies are those that disappear. They weave themselves into the fabric of everyday life until they are indistinguishable from it}}. The idea of disappearance, here, is as much physical as cognitive.
\smallskip

\noindent From the mid-1990s onwards, this vision was technologically feasible thanks to the advent of mobile telephony and the progress it had made possible in terms of miniaturization of electronic components and wireless communication. At MIT (Massachusetts Institute of Technology), Hiroshi Ishii, a researcher at the Media-lab, is already wondering about the consequences of such an evolution (\textit{objects that think}, \cite{hawley1997things}). In 1998, largely inspired by Mark Weiser's work, Eli Zelkha introduced the concept of \textit{Ambient Intelligence}  (AmI) \cite{zelkha1998devices} which aims to bring together the computer resources embedded on everyday objects to assist people in their daily activities. In 1999, Kevin Ashton adopted the term \textit{Internet of Things} (IoT) \cite{ashton2009internet} which reflects the integration of the physical world into the Internet through sensors spread throughout our environments. In a broad sense, the notion of \say{things} (e.g., house, building, garden, etc.) goes well beyond that of \say{objects} (e.g., chair, watch, lamp, etc.). The concept of \textit{connected object} then appeared in order to characterize the everyday objects connected to this network.
\smallskip

\noindent In the early 2000s, the term \textit{Pervasive Computing} was introduced. Taking up the key ideas of ubiquitous computing as stated by Mark Weiser, both concepts are used interchangeably \cite{satyanarayanan2001pervasive}. In 2002, the notion of \say{\textit{Disappearing Computing}}\footnote{\url{http://www.disappearing-computer.org}}, very similar to that of \say{\textit{Calm technology}}, appeared as part of an initiative of the European Commission (Future and Emerging Technologies, FET).
\smallskip

\noindent In 2006, Helen Gill (National Science Foundation) introduced the term \textit{Cyber-Physical Systems} (CPS) \cite{lee2016introduction} which generalizes the concept of embedded systems to that of connected objects with the objective of coordinating them with physical processes \cite{rajkumar2010cyber}. Since 2007, CPSs have been a national priority in the United States\footnote{\url{https://www.nsf.gov/geo/geo-data-policies/pcast-nit-final.pdf}}. Just as the Internet has profoundly transformed the way humans interact with each other, CPSs will transform the way they interact with the physical world \cite{rajkumar2010cyber}.\\
\noindent In 2008, mainly due to the explosion of mobile devices (phones, tablets, etc.), there are more connected objects than people on earth\footnote{\url{https://www.cisco.com/c/dam/en_us/about/ac79/docs/innov/IoT_IBSG_0411FINAL.pdf}}. Dominique Guinard laid the foundations of the \textit{Web of Things} (WoT) \cite{guinard2009towards} which integrates the objects at the application layer of the OSI model. It is then possible to use a common interface to access objects (e.g., RESTful web services, Constrained Application Protocol (CoAP), Message Queue Telemetry Protocol (MQTT), etc.), to compose applications based on connected objects, etc.
\smallskip

\noindent Since then, many concepts have emerged, often promoted by industries or even states that have understood the eminently strategic dimension of these technologies. In particular, in the early 2010's, Cisco\textsuperscript{\texttrademark} introduced the concept of the \textit{Internet of Everything} (IoE) which, beyond Machine-to-Machine (M2M) communications, encompasses technology assisted People-to-People (P2P) communications and Machine-to-People (M2P) communications (via, for instance, social networks or other digital media). In line with CPS, the concepts of \textit{Industrial Internet of Things} (IIoT) \cite{BOYES20181} and \textit{Industry 4.0} \cite{kagermann2013recommendations}, by their highly strategic dimension, are gaining some momentum and, within the European Union, widely developed in Germany.
\medskip

\noindent This brief historical overview highlights a number of concepts that characterize the fusion of the physical and digital worlds, prophesied by Mark Weiser as early as 1991. So as to clarify the picture, we provide, in the first part of this paper, a synthesis of the main concepts (\S\ref{synthesis}). Without claiming to be exhaustive, this synthesis makes it possible to clarify their particularities and overlaps in terms of application and scientific research areas.
\medskip

\noindent It results from the synthesis (\S\ref{results}) that the information processing systems underlying these concepts have in common that their purposes are achieved in the physical environment through actions effecting some of its properties. This common denominator brings them together under the term \textit{ambient systems}. This semantics is justified by the general system theory \cite{von1968general} which defines an ambient environment by all the external factors which, by acting on certain properties of a physical system, determine its evolution.
\medskip

\noindent By interacting with the physical environment, these systems inherit its complexity. This complexity is also driven by their infrastructure and the rationality of their decisions. These elements are detailed in \S\ref{complexity}. They justify an observation resulting from the synthesis : since their effects are unpredictable, (1) responsibility for actions is often left to users, (2) otherwise, there are risks to individuals and their environment. Mark Weiser's vision of disappearing computing is still far from being a reality. 
\smallskip

\noindent This situation calls for an epistemological rupture that we propose to concretize through the systemic approach whose methodology is presented in \S\ref{epistemic}. Without claiming to predict, the proposed approach makes it possible to evaluate \textit{in vivo} the \textit{effectiveness} of these systems.
\smallskip

\noindent To conclude (\S\ref{conclusion}), we introduce some perspectives that the evaluation of the effectiveness can bring to these systems.

\section{Thematic classification}
\label{synthesis}
\noindent It is worth setting up a methodological framework which, without claiming for exhaustiveness, makes it possible to study the particularities and the commonalities of the main concepts highlighted in terms of application and research areas, namely \say{Internet of Things}, \say{Ubiquitous Computing}, \say{Pervasive Computing}, \say{Ambient Intelligence} and \say{Cyber-Physical Systems}. Figure \ref{evolution} shows the evolution of the number of scientific publications relating to them. The graph was compiled from Scopus\textsuperscript{\textregistered}, a database of citations and abstracts from peer-reviewed scientific publications (journals, books and conferences). The results obtained cannot be exhaustive here, as other databases are available (e.g., IEEE, ACM, Science Direct, etc.). The objective, while not exhaustive, is to identify a trend.

\input{applications.tex}
\begin{table*}
  \small 
  \begin{tabular}{m{0.85\textwidth}m{0.1\textwidth}}
  \hline
 \textbf{Scopus\textsuperscript{\textregistered} queries} & \textbf{\#publications} \\
  \hline
  \rowcolor[HTML]{EFEFEF} 
  \raggedright(KEY("internet of things") OR KEY("internet of thing") OR KEY(iot)) AND (LIMIT-TO(SUBJAREA,"COMP")) AND (LIMIT-TO(LANGUAGE,"English")) & 42967 \\
  \raggedright KEY("ubiquitous computing") AND (LIMIT-TO(SUBJAREA,"COMP")) AND (LIMIT-TO(LANGUAGE,"English")) &  30047 \\
  \rowcolor[HTML]{EFEFEF} 
  \raggedright (KEY("cyber physical system") OR KEY("cyber physical systems") OR KEY("cyber-physical system") OR KEY("cyber-physical systems") OR KEY("cps")) AND (LIMIT-TO(SUBJAREA,"COMP")) AND (LIMIT-TO(LANGUAGE,"English")) & 9822 \\
  \raggedright KEY("pervasive computing") AND (LIMIT-TO(SUBJAREA,"COMP")) AND (LIMIT-TO(LANGUAGE,"English")) &  4457 \\
  \rowcolor[HTML]{EFEFEF} 
  \raggedright KEY("ambient intelligence") AND (LIMIT-TO(SUBJAREA,"COMP")) AND (LIMIT-TO(LANGUAGE,"English")) &  3061 \\
  \hline
  \end{tabular}
  \caption{\label{queries} Scopus\textsuperscript{\textregistered} queries for keywords extraction.}
  \end{table*}

\input{chronologie.tex} 
\input{overview.tex}

\noindent The methodology chosen consists in determining all the keywords associated with articles published over the period 1989-2019 relating to the various concepts, and classifying them according to their frequency of appearance (1989 being the year from which the term Ubiquitous Computing was introduced). Scopus{\textsuperscript{\textregistered} queries are given in the table \ref{queries}.
\smallskip

\noindent In the sequel, the different concepts are analysed on the basis of these keywords (Denoted by \kwrd{keyword}). The figure \ref{apps} depicts the main application domains associated with each of the concepts. The figure \ref{cartography} shows a synthesis of the most common keywords associated with the concepts used in the study. This figure also depicts the relationships that can exist between the different concepts. In the sequel, these results serve as the basis for the analysis of each of the concepts.

\subsection{Internet of Things}
\noindent In its common sense, the notion of object is associated with any material entity perceived in our environments and assigned to a specific usage (e.g., table, book, chair, etc.). Recent advances in electronics (e.g., reduction of power consumption, miniaturization of components, wireless communication technologies, etc.) make it possible to envisage new usages based on the digital exploitation of contextualized data resulting from the interactions of these objects with the environment. 
\smallskip

\noindent Objects are equipped with \textit{resources} allowing to process \textit{digital data}. This processing makes it possible, on the one hand, to translate this data into different forms of energy through actuator/effector chains. It also makes it possible to convert different forms of energy into numerical data using sensors. The use of these resources leads to a multidimensional variability in the ability of the objects to perceive their environment and, through their energy activity, to be perceived by both humans and machines. The perceptible modification of certain environmental properties (e.g., radio waves, electromagnetic waves, etc.) then allows objects to communicate. The organized set of these elements with which the objects are equipped forms a \textit{device}.
\smallskip

\noindent On this basis, it becomes possible to connect objects to a network through OSI\footnote{Open Systems Interconnection} hardware layers. The Internet of Things has been defined by the International Telecommunication Union (ITU) as \say{\textit{a global infrastructure for the information society, enabling advanced services by interconnecting (physical and virtual) things based on existing and evolving interoperable information and communication technologies (ICT)}}.
\smallskip

\noindent In this context, new forms of interaction emerge from these devices, embodied through remotely accessible software services, resulting in the fusion of the digital and physical dimensions of our environments. Thereby, we define herebelow the notions of \textit{connected thing} and \textit{connected object} as follows:
\smallskip

\noindent \textbf{Connected thing} -- \textit{A material entity made useful to individuals by means of organized electronic and computer resources that, together with the material entity, form a finalized system, technological artifact formally accessible and identifiable within a computer network.}
\smallskip

\noindent The notion of connected thing then generalizes that of connected object by integrating, for example, entities delimited in space (building, house, factory, garden, forest, etc.). A thing does not necessarily have an identified primary function as it is for an object.
\smallskip

\noindent \textbf{Connected object} -- \textit{An everyday object whose primary function is transcended by organized electronic and computer resources that, together with the object, form a finalized system, technological artifact formally accessible and identifiable within a computer network.}
\smallskip

\noindent With these elements in mind, the structural model of a connected thing is described in figure \ref{object}. This model is based on those described in \cite{de2012internet} and \cite{waldner2007nano}.

\input{object.tex}
\smallskip

\noindent \textbf{The extraction of the keywords associated with scientific publications shows three predominant research areas related to the Internet of Things}:

\begin{enumerate}
  \setlength\itemsep{0.0em}
  \item  The first research area focuses on \textbf{technological aspects}, both from the point of view of the devices \textbf{that are highly constrained} (\kwrd{Low Power Electronics}) and \textbf{wireless communication protocols} (\kwrd{WLAN}, \kwrd{Zigbee}, \kwrd{Wi-Fi}, \kwrd{Bluetooth}, \kwrd{5G}, etc.) and network (\kwrd{Next Generation Neworks}, \kwrd{Gateways}, \kwrd{Routing Protocols}, etc.). The current diversity of protocols is at the heart of interoperability issues (\kwrd{Interoperability}) that ETSI (European Telecommunications Standards Institute) in Europe, together with other standardisation actors around the world, is trying to solve (OneM2M \cite{swetina2014toward}).
  \item The second research area concerns \textbf{sensor data}. Due to their volume (\kwrd{Big Data}), these data require special storage and processing strategies (\kwrd{Edge}, \kwrd{Fog Computing}). 
  \item Finally, the research focuses on \textbf{security aspects}, particularly those related to data (\kwrd{Security \& Privacy}, \kwrd{Mobile Security}, \kwrd{Blockchain}, \kwrd{Trusted Computing}, etc.).
  \end{enumerate}

 \noindent \textbf{From an application point of view, three areas emerge: smart-cities/homes/buildings, healthcare and energy, mainly oriented towards monitoring}:
 \begin{enumerate}
  \setlength\itemsep{0.0em}
  \item \textbf{In so-called smart cities} (\kwrd{Smart City} \cite{minoli2018internet}), sensors provide information allowing to effectively manage resources and assets. These range from optimised road traffic management \cite{masek2016harmonized}, transportation \cite{santi2014quantifying}, parking spaces \cite{khanna2016iot}, lighting \cite{castro2013smart}, to water supply network \cite{dickey2018smart}, waste management \cite{medvedev2015waste}, etc.\\  
  \noindent For instance, since 2006 in Paris, RFID chips have been installed in the city's 95,000 trees, allowing them to be accurately monitored and maintained. Per Caroline Lohou, head of the parks, gardens and green spaces department at the Paris city hall, \say{\textit{the logger has one click access to everything he needs to treat the trees: watering, pruning, fertilization, treatments, old address if the tree has been transplanted, etc.}}.

\item \textbf{In the healthcare area} (\kwrd{Healthcare} \cite{baker2017internet}), IoT's advantages are also undeniable. In New York, the Mt. Sinai Medical Center, in partnership with GE Healthcare, has implemented an intelligent bed allocation management that has resulted in a 50\% reduction in patient referrals time \cite{thomas2013automated}. On their side, Stanley Healthcare\footnote{\url{https://www.stanleyhealthcare.com}} offers hospitals a real-time geolocation (\kwrd{Location}) solution for their patients, staff and medical devices. Thereby, hospitals can thus optimize management and improve patient care. Via connected pill boxes, the various actors in the care pathway are able to know whether or not a patient has taken the prescribed treatment (therapeutic compliance) \cite{benhamou2018observance}. Even more reliable, the connected absorbable pill (Abilify MyCite from the Otsuka laboratory), available in the United States, allows therapeutic compliance for patients in psychiatric hospitals. Finally, deformable/stretchable electronic patches that adhere to the skin and track body movements provide doctors with real-time information on patients' health status (epidermal electronics \cite{tamura2018seamless}).
\item Finally, \textbf{in the field of energy} (\kwrd{Energy Efficiency}, \kwrd{Energy Utilization}, \kwrd{Smart Power Grids}), IoT makes it possible to support the energy transition by complying, for instance, with the 2009 European directive on the smart grids deployment \cite{ue2009}. Thus, by 2021, more than 35 million Linky electricity consumption meters will be installed in France. The meter is equipped with sensors and uses the existing network to receive and transmit data without the physical intervention of a technician.
 \end{enumerate} 

 \begin{mdframed}
  \noindent Connected things, through sensors, allow the acquisition of data relating to humans and their environment. However, the valuation of this data is often the responsibility of users. Thus, the treatment of trees in the city of Paris remains the responsibility of the loggers; the optimization of patient care in hospitals to that of the nursing staff, therapeutic compliance to that of the patients, etc.
 \end{mdframed}

\subsection{Ubiquitous Computing}
\noindent Ubiquitous computing, by moving computational resources from the personal computer to everyday objects, \textbf{invites researchers to rethink the ways in which humans and computers interact}.
\smallskip

\noindent This research area (\kwrd{Human Computer Interaction}) is predominant, as is that concerning mobile devices (\kwrd{Mobile Devices}, \kwrd{Telephone Sets}, \kwrd{Mobile Phones}, \kwrd{Wearable Computers}, etc.), vectors of these new interaction modalities through services with high added-value (\kwrd{Location Based Services}, \kwrd{Multimedia Services}, \kwrd{Visualization}, etc.). For instance, connected glasses (\kwrd{Wearable Computers}), often associated with Eyewear Computing \cite{tag2018eyewear} are a representative example of the new modalities of interaction with the computer \cite{bace2016ubigaze}. In the field of maintenance, the relevance of these new interaction modalities is all the greater because it allows significant time and efficiency gains thanks to information elements superimposed on reality (\kwrd{Augmented Reality}) \cite{champalle2008ordinateur}.
\smallskip

\noindent \textbf{From an application point of view, two areas emerge from the study of keywords: healthcare and smart environments (intelligent buildings, smart-homes and smart-buildings)}:
\begin{enumerate}
  \setlength\itemsep{0.0em}
      \item \textbf{In the field of healthcare} (\kwrd{Healthcare}) the most emblematic device is the connected watch (\kwrd{Wearable Computers}) which, based on a certain number of physiological data, makes it possible to prevent cardiovascular or cerebral risks, to detect falls \cite{paret2019wearables}, biological dangers \cite{miaolei2019chemically}, etc. The LVL bracelet from the American company BSX measures the wearer's cardiac parameters and hydration level. It signals in real time the amount of water to be absorbed via a dedicated application on the smartphone.
      
    \item \textbf{In the field of so-called intelligent buildings and houses} (\kwrd{Smart Home}, kwrd{Intelligent Buildings}), sensors embedded on clothing pave the way for new interactions. For instance, biometric sensors sewn into clothing allow temperature and lighting to be modulated in a contextual way (\kwrd{Context Aware Computing}, \kwrd{Location}) for well-being and energy saving purposes (\kwrd{Energy efficiency},\kwrd{ Energy Utilization}) \cite{essa2000ubiquitous}\cite{ojuroye2017smart}.\\
    The Homni \cite{homni} connected lamp, developed by Teraillon, adjusts its brightness throughout the day. Equipped with sensors for room temperature, brightness, humidity and noise, it also makes it possible to offer diagnostics and advice via the dedicated application \say{Terraillon Wellness Coach} to improve sleep quality. It also includes a bluetooth speaker allowing to broadcast a suitable sound atmosphere in the early and late hours of the night.\\
    Connected switches (e.g., the SmartPeeble portable switch/variator from Awox) and connected locks complete the possibilities offered by connected objects within the home (e.g., Smart Lock from the Austrian company Nuki is a box that can be installed on a lock and allows it to be opened and closed remotely via a dedicated application). Another example is the refrigerator, a connected object that is emblematic both for the fact that it is one of the first objects to have been connected (LG, 2000) and for the usage scenarios it suggests. Using RFID chips attached to food, the refrigerator reads its contents, offers adapted menus and warns users when food is missing or outdated \cite{luo2008smart}\cite{luo2009smart}.\\
    Finally, the Luminon of the Ubiant startup is a watchtower that connects to all connected objects in the home and indicates by its colour (green or red) whether household consumption exceeds the predefined target. If the consumption target is exceeded, the user receives a notification on his smartphone to warn him of the location of the excess consumption.
\end{enumerate}

\begin{mdframed}
  \noindent Ubiquitous computing offers humans, via mobile devices, new ways of interacting with computer. However, it must be noted that these interactions most often result in a set of notifications from applications associated with the devices and are all the more restrictive because they require the permanent availability of users.
\end{mdframed}

\subsection{Pervasive Computing}
\noindent \textbf{Pervasive computing has strong connections with ubiquitous computing}. This is very clear from the figure \ref{cartography}. However, the predominant research area here is on middleware (\kwrd{Middleware}), an architectural approach to mediation and orchestration of services \cite{footen2012service}, some of which being executed on connected things.
\smallskip

\noindent The exchange of messages (e.g., Message Queue Telemetrie Transport, MQTT) and the implementation of \textit{remote software services} (e.g., Service Oriented Approach, SOA) allow computer applications to interact, cooperate and exchange information with connected things (System of Systems). The dynamic nature of the everyday objects and services they carry means that they can appear and disappear over time. The implementation of \textbf{service discovery protocols} (\kwrd{Service Discovery}) is an important research area that allows middleware to be aware of their execution context (\kwrd{Context Aware Services}), a context that is defined by the purposes of the system concerned \cite{bazire2005understanding}.
\smallskip

\noindent \textbf{From an application point of view, the same domains as above emerge with a strong predominance for the healthcare domain (\kwrd{Healthcare})}.
\smallskip

\noindent There are many examples of middleware implementation \cite{becker2019pervasive}. Context consideration is characterized by \textit{adaptive systems} (\kwrd{Adaptive Systems}) that can rely on semantic annotations (\kwrd{Semantics}) to select the most relevant available services available \cite{muhammad2018review} to ensure continuity and \textit{quality of service} (\kwrd{Quality of Service}, QoS) \cite{masip2018managing}.

\subsection{Ambient Intelligence}
\noindent Ambient intelligence, as defined by Eli Zelkha, \textbf{aims to bring together the computer resources embedded in objects to assist people in their daily activities} (\kwrd{Activities of Daily Life}).
\smallskip

\noindent The elderly (\kwrd{Elderly People}), often dependent because they are restricted in their freedom of movement and are subject to a decline in their cognitive abilities, are the ideal target of ambient intelligence. The economic and social impacts that ambient intelligence promises are enormous in this context \cite{sun2009promised} both on the prospects of reducing the costs of their care by society and their families and on improving their quality of life (\kwrd{Quality Of Life}).
\smallskip

\noindent \textbf{Two research areas emerge clearly from the study of keywords related to ambient intelligence} (figure \ref{cartography}):
\begin{enumerate}
  \setlength\itemsep{-0.1em}
  \item \textbf{Research on artificial intelligence} to assist people in their daily activities (\kwrd{Ambient Assisted Living}, AAL). Addressing the cognitive and physical deficiencies of these people suggests a form of autonomous intelligence (\kwrd{Intelligent Systems},\kwrd{Autonomous Agents}) that can interpret the needs of these people (\kwrd{Activity Recognition}, \kwrd{Gesture Recognition}, \kwrd{Speech Recognition}) and make relevant decisions for them (\kwrd{Decision}, \kwrd{Support Systems}, \kwrd{Decision Making}, \kwrd{Knowledge Based Systems}).
  \item \textbf{Research on the robotization of environments} for support and physical assistance to people (\kwrd{Intelligent Robots}, \kwrd{Home Automation}.)
\end{enumerate}
\smallskip

\noindent \textbf{From an application point of view, people assistance (\kwrd{Ambient Assisted Living}) is logically the main field of application with repercussions in the fields of robotics, smart environments (\kwrd{Intelligent Buildings}, \kwrd{Smart Home}) and healthcare (\kwrd{Healthcare}) \cite{rashidi2013survey}\cite{pieper2011ambient}}.\\
\noindent There are many examples and usage scenarios. For example, through the means of communication available in the home (televisions, tablets, lamps, speakers, etc.) an intelligent system can send messages that are useful to people (e.g., \say{\textit{think about taking the medication}}) and adapted to their physical condition (e.g., deafness, visual impairment).\\
\noindent Another example concerns the use of sensors scattered throughout the habitat from which an intelligent system is likely to recognize people's activity in order to anticipate their needs or prevent rescue in the event of a fall or abnormal activity suggesting a physical problem.\\
\noindent Combined with activity recognition, connected objects such as lamps, blinds, loudspeakers, radiators, etc. allow intelligent systems to learn people's behavioural habits in order to improve their well-being by reproducing their favourite visual and sound environments (\kwrd{Decision Making}) or by suggesting relevant actions (\kwrd{Decision Support Systems}). A concrete example is the Nest thermostat, which controls the boiler by observing the habits of the occupants of the house in order to optimize energy consumption \cite{hernandez2014smart}.
\medskip

\begin{mdframed}
  \noindent Ambient intelligence, by suggesting a form of robotization of our habitats, concretizes the need for assistance to people, especially those whose cognitive and/or physical abilities are declining. However, the promised assistance is most often translated into a set of notifications that require the permanent availability of users or their carers.
\end{mdframed}

\subsection{Cyber-Physical systems}
\noindent Cyber-physical systems (CPSs) are a \textbf{generalization of the concept of embedded systems to that of connected things with the objective of making them collaborate for the control of physical processes} \cite{rajkumar2010cyber}. The purposes of these systems are achieved through the physical environment (\kwrd{Physical Systems}, \kwrd{Physical Environments}, \kwrd{Physical World}) autonomously (\kwrd{Robots}, \kwrd{Robotics}, \kwrd{Actuators}).
\smallskip

\noindent \textbf{Two research areas emerge from the study of keywords related to these systems (figure \ref{cartography}) which could be related to \say{Trustworthy Systems of Systems} \cite{steinhogl2015trustworthy})}:
\begin{enumerate}
  \setlength\itemsep{0.0em}
  \item Research that addresses \textbf{cyber security aspects} (\kwrd{Network}, \kwrd{Security}, \kwrd{Intrusion Detection},\\ \kwrd{Cyber-Attack}, \kwrd{Cyber Security}, \kwrd{Security Systems}),
  \item Research related to \textbf{risks and their management} (\kwrd{Risk Assessment}, \kwrd{Accident}, \kwrd{Prevention}). From this point of view, it is interesting to note the efforts regarding the formal verification of these systems (\kwrd{Formal Verification}). 
\end{enumerate}
\medskip

\noindent\textbf{From an application point of view, three areas emerge: energy, industry and robotics}:
\begin{enumerate}
  \setlength\itemsep{0.0em}
      \item \textbf{In the fields of energy and smart power grids} (\kwrd{Smart Power Grids}) \cite{keyhani2016design}, the aim is to adjust electricity production and distribution according to real time demand in a targeted manner. For instance, Linky meters allow to determine electricity demand in real time. Depending on this demand, local production resources favouring renewable energy sources (solar panels, wind power) are used to meet this demand. In the long term, CPS should allow the advent of Zero-Net Buildings (ZNEs) \cite{krarti2016evaluation} that, using sensors, solar panels, LED lighting and batteries, do not use more energy than they produce,
      \item \textbf{In the field of industry} (\kwrd{Industry 4.0}, \kwrd{Manufacturing}) \cite{alcacer2019scanning}, a striking example is that of stock management at Amazon, carried out by robots that autonomously walk the warehouse and constitute orders using tags present on each of the products \cite{correll2018analysis},
      \item Finally, the implementation of robots such as \textbf{autonomous vehicles} (\kwrd{Vehicles}) promises to improve the fluidity of road traffic. Connected to online services and equipped with sensors, their perception and reactivity, by not being subject to fatigue or inattention, are much better than that of humans. These vehicles are therefore intended to drastically reduce accidents due to fatigue, alcohol, etc., (94\% of which being of human origin \cite{autonomouscars}), and costs by increased driving efficiency.
\end{enumerate}
\medskip

\begin{mdframed}
  \noindent CPSs embody a robotization of the world whose interest lies mainly in the optimization of resources, their means of supply, etc. However, by interacting with their environment, these systems are not without risks, as evidenced by the current lines of research concerning them.
\end{mdframed}

\subsection{Synthesis}
\label{results}
\noindent This brief synthesis highlights the fact that \textbf{purpose of the systems underlying the different concepts discussed, beyond the collection of environmental data from sensors, is mainly the implementation of relevant actions that the processing of these data suggests in our environments}:
\begin{enumerate}
  \setlength\itemsep{0.0em}
  \item In cities, it is a question of controlling resources and assets in order, for example, to fluidify traffic \cite{masek2016harmonized}, to manage energy distribution (water and electricity) \cite{ue2009}, etc.,
  \item Within the home, it is a question of controlling electrical appliances in order to optimize household consumption \cite{hernandez2014smart}; modifying certain physical parameters in order to assist and improve people's well-being \cite{homni}, their safety \cite{viderberg2019security}, etc.,
  \item At the human level, it is about influencing behaviours to improve physical and cognitive performances \cite{benhamou2018observance}.
\end{enumerate}
\medskip

\noindent This purpose, common denominator of the concepts discussed, brings them together under the term \textit{ambient systems}. This semantics is relevant and reflects well the idea of the fusion of the physical and digital worlds. It finds roots in the general system theory \cite{von1968general} which defines an ambient environment by all the external factors which, by acting on certain properties of a physical system, determine its evolution.

\noindent Nevertheless, responsibility for actions is still left to individuals, these systems only providing support for decision-making (\kwrd{Decision Support Systems}). This is evidenced by the mobile applications associated with these systems, most of which being comparable to remote controls. However, the multiplication of dedicated applications and the notifications resulting from their analyses induce a cognitive overload which has the consequences of reducing users' performance \cite{leroy2009so} and inducing stress \cite{mark2008cost} likely to motivate the abandonment of the systems and their applications \cite{pielot2017beyond}\cite{shih2015use}. Not to mention the risks to both humans and their environment that the decisions made by these systems are likely to cause. This situation is explained by the complexity of these systems \cite{coutaz2008plan}. This complexity, by preventing the accurate modelling of these systems, also makes it impossible to predict with certainty the effects that these systems are likely to have in the environment.
\smallskip

\noindent The next paragraph is devoted to formalizing this problem and justifying the complex nature of these systems.

\section{Complex systems, irreducible to a model}
\label{complexity}
\noindent A definition of the notion of system appeared in the 1940s as part of the work on the systemic approach, which is at the origin of currents of thought such as structuralism in France or cybernetics in the United States. From the 1970s, under the influence of Ludwig von Bertalanffy \cite{von1968general}, this work was extended to \textit{emerging systems} and the concept of \textit{self-organization}. In this context, a system is defined by: 
\smallskip

\noindent \textbf{System} -- \textit{A system is a set, forming a coherent and autonomous unit, of real or conceptual objects (material elements, individuals, actions, etc.) organized according to a goal (or a set of goals, objectives, purposes, projects, etc.) by means of a set of relationships (mutual interrelationships, dynamic interactions, etc.), all immersed in an environment} \cite{le1992systemique}.
\medskip

\noindent The introduction of autonomy in this definition refers to the intrinsic capacity of a system to maintain, by itself, its unity in order to perpetuate the achievement of its purposes. The action of maintaining suggests feedback loops, a particular form of interaction whose study is at the heart of cybernetics: the system modifies the environment which, in turn, modifies the system within circular relationships.

\noindent As an integral part of these systems, computer systems (or, more generally, digital systems) can be defined as follows:
\smallskip

\noindent \textbf{Computer system} -- \textit{A computer system is a set of computer, telecommunications equipment and software used to collect, process, store, transmit and present data} \cite{lonchamp2017introduction}. 
\medskip

\noindent It is a technological artifact whose purposes are achieved through \textbf{data manipulation}. All the IT and telecommunications resources form the coherent unit that refers here to the notion of infrastructure, which will have to be maintained in line with the purposes of the system. For instance, mechanisms for periodically refreshing the contents of DRAMs (Dynamic Random Access Memory) or hardware redundancy ensure the integrity of the data stored in the memory. Cyclic redundancy (CRC) or Error-Correcting Code (ECC) control mechanisms ensure the integrity of data as it is transmitted over a network. Consequently, the achievement of the purposes of these systems is \textit{deterministic}, being conditioned only by the validity of the means of data processing implemented, regardless of time and space. This determinism allows the implementation of models and tests that allow the validation and prediction of the results of data processing.
\smallskip

\noindent An ambient system, fusion of the digital and physical worlds, is a system of systems whose singularity lies in the fact that its purposes are achieved, no longer exclusively through the manipulation of data, but also through the manipulation of some properties of the physical environment. \textbf{It is no longer merely a question of collecting, processing and storing data, but of \textit{observing}, \textit{modifying} and \textit{maintaining} properties of the physical environment} through transducers (sensors and actuators), interfaces between the physical and digital worlds at the heart of their infrastructure.\\
In fact, the realization of the purposes of an ambient system is part of a set of dimensions (including spatio-temporal dimensions) that characterize the notion of \textit{context}. For Bazire et Brézillon \cite{bazire2005understanding}, the context is defined by the purposes of the system concerned: \say{\textit{context acts as a set of constraints that influence the behavior of a system dedicated to a certain task}}. 
\smallskip

\noindent The set of constraints here is threefold:
\begin{itemize}
  \setlength\itemsep{0.0em}
  \item[--] First, it corresponds to the set of external factors inherent to the \textbf{physical environment}, complex system in non-linear interactions with the ambient system and possibly in conflict with the effects it produces,
  \item[--] It then corresponds to the \textbf{latent dynamicity of the infrastructure} inherent to the connected objects that do not offer any guarantee of interoperability and spatio-temporal availability,
  \item[--] Finally, it corresponds to the \textbf{rationality of the decisions} specific to the decision-making algorithms underlying the achievement of their purposes through transducers embedded on connected things (in this context, the notions of substantive and procedural rationalities introduced by Herbert Simon are particularly relevant). 
\end{itemize}  
\smallskip

\noindent These constraints, which are as much challenges in maintaining the unity of ambient systems and in achieving their purposes, are justified and detailed below.

\subsection{The challenge of the physical environment}
\noindent The purposes of ambient systems are achieved through the physical environment. In other words:
\begin{itemize}
  \setlength\itemsep{0.0em}
  \item[--] It is no longer a question of storing a value in memory whose persistence is guaranteed by design, but of maintaining a set of properties of the physical environment around a point of equilibrium (the temperature of a room for example), which many physical processes, independently of the ambient system, are likely to disturb in a random manner,
  \item[--] It is no longer a question of transmitting information by bounded or guided channels (optical fibre, coaxial cables, etc.), but by air channels (wireless telecommunications), the keystones of ambient systems whose sensitivity to disturbances inherent to the physical environment has been demonstrated. In particular, temperature is a factor in the degradation of data transmission performance \cite{boano2010impact}\cite{marfievici2013environmental},
  \item[--] It is not a question of connecting the components of ambient systems to the wired electrical network, but of collecting and storing energy. The technological constraints associated with batteries and weather variations are potential disruptive factors here,
  \item[--] Finally, it is no longer merely a question of \textit{controlled systems}, reducible to a model, for processing and transforming information but, more broadly, of complex \textit{open systems}, irreducible to a model, leading irreversible transformations of physical properties \cite{le1990modelisation}. 
\end{itemize}
\smallskip

\noindent Before going further, it is necessary to introduce the notions of entropy, negentropy, controlled systems and open systems.
\smallskip

\noindent The notion of entropy was introduced by Rudolf Clausius in 1865 and is at the heart of the second principle of thermodynamics; it has since been adopted in many fields including statistical physics, information theory, mathematics, philosophy, human sciences, etc.\\
\noindent In the sense of thermodynamics, entropy characterizes the degree of dispersion, the homogenization of the energy of a system that no longer allows it to produce organized effects. Thus, entropy characterizes disorder, the disorganization of the system.
\noindent In the sense of statistical physics, the entropy of a system characterizes the lack of information relating to the organization of its constituents. Entropy is minimal if the macroscopic state of the system is the representation of a single organization of its constituents. It increases with the number of possible organizations that lead to the same macroscopic state. In other words, the higher the entropy, the less informative the macroscopic state is about the actual internal organization of the system.
\smallskip

\noindent For instance, a macroscopic state for a set of 52 playing cards may correspond to \say{\textit{all cards are hidden}}. Entropy is minimal, this macroscopic state is indeed the representation of a single organization of the cards.
  \noindent When the macroscopic state corresponds to \say{\textit{all cards are hidden except one}} (the position of the visible card being important), entropy increases (the macroscopic state is the representation of possibly $52^2=2704$ organizations). Indeed, the macroscopic state of the system is no longer the representation of a single organization of its elements. It is not informative of the underlying real organization.
\noindent Entropy is maximum when the macroscopic state corresponds to \say{\textit{all cards are visible}} (the macroscopic state is the representation of possibly $52!=8.06\times10^{67}$ organizations).
\smallskip

\noindent The notion of representation calls for that of \textit{modelling} which, in its simplifying logic, is subject to abstractions that cannot allow it to be a single representation of the real organization of the constituent of a system. On this basis, a definition of the entropy of a system could then be as follows:
\smallskip

\noindent\textbf{Entropy of a system} -- \textit{The entropy of a system characterizes the degree of ignorance, embodied in the impossibility of a model to predict the real evolution of the elements of the system. The higher the entropy of a system, the less its model is reliable to predict its future evolution.}
\medskip

\noindent The degree of ignorance is highlighted in the \textit{principle of maximum entropy}. When the (imperfect) knowledge of a phenomenon is represented by a probability law, the principle of maximum entropy consists in choosing, among the set of distributions responding to the constraints specific to the phenomenon, the one with the greatest entropy, i.e., the one with the highest variance and which is therefore the least informative.
\smallskip

\noindent In light of entropy, the definition of a system can be completed in different ways. In the context of systems whose purposes are achieved through the physical environment, onre may refer to so-called open or controlled systems. 
\smallskip

\noindent\textbf{Open system} -- \textit{An open system is in permanent relationship with its environment with which it exchanges energy, matter and information. The entropy of the system can then increase, stabilize or decrease as these reorganizing exchanges progress (figure \ref{systems}.a). Entropy is minimal if the evolution of the system, depending on these exchanges, remains predictable (cf. negentropy).}
\medskip

\noindent\textbf{Controlled system} -- \textit{A controlled system is a system in which all the exchanges of energy, matter and information likely to transform its organization are regulated by feedback loops in active opposition to the increase in its entropy, i.e., constrains the evolution of the system so that it remains predictable (figure \ref{systems}.b).}
\medskip

\noindent Controlled systems are said \textit{neguentropic} systems \cite{morin2013methode} 
\medskip

\noindent\textbf{Negentropy} -- \textit{Negentropy, unlike entropy, characterizes the degree of knowledge available to enable a system to overcome entropy, i.e., to maintain the predictability of its future evolutions. The entropy of a system is minimal if its neguentropy is maximal.}

\input{systems.tex}

\noindent Ambient systems often use feedback loops to counter entropy. In particular, ambient systems \textit{auto-adaptive} (\kwrd{Adaptive Systems}) which, through observations (\kwrd{Adaptive Systems}), adapt to changes through feedback loops (e. g. Models@runtime, MAPE-K (Monitor-Analyze-Plan-Execution over a shared Knowledge) as part of Autonomic Computing).
\smallskip

\noindent Nevertheless, while all the exchanges of energy, matter and information that contribute to the preservation of the organization of a system can, under certain conditions, be known and controlled (like cyber-physical systems in an industrial environment), it is illusory to think that they can be completely known within the physical environment. \textbf{In other words, except in special cases, it is unrealistic to expect to control all the sources of entropy of a system whose objectives are achieved through the physical environment (figure \ref{systems}.c), irreducible to a model and therefore complex}.
\medskip

\noindent Beyond the physical environment in which they operate, the complexity of ambient systems is also related to the \textbf{dynamicity and variability of their infrastructure}, inherent in a set of contextual, technological and environmental factors that are detailed in the following paragraph.

\subsection{The challenge of the infrastructure}
\noindent Infrastructure is at the heart of the ambient systems organization. It refers to the notion of (structural) unity that these systems must maintain in order to perpetuate the achievement of their purposes. It is characterized by a set of technological means (such as those discussed in the context of connected things, i.e., \kwrd{Devices}, \kwrd{Resources}, \kwrd{Services} and \kwrd{Energy}), immersed in the physical environment and interrelated within a communication network (\kwrd{Network}). \textbf{Thereby, the infrastructure of ambient systems is subject to dynamics and variability that may call into question its adequacy with the purposes of the system.}
\smallskip

\noindent The main factors leading infrastructure dynamics and variability are technological, contextual and environmental.

\subsubsection{Contextual factors}
\noindent When the purpose of an ambient system is to maintain a property of the physical environment at a certain value, this purpose only makes sense if it is spatially bounded, i.e., concerns a well-defined subspace of that environment (e.g., a kitchen, a building, etc.). It is therefore clear that the purposes of an ambient system can only be achieved in presence of relevant devices and services, i.e. capable of influencing the physical property in question within the spatial boundaries of the subspace concerned (figure \ref{contextual}. a)). However, even if this were the case: (1) the nomadic nature of the connected objects does not guarantee the availability in time and  space of their associated devices and services (figure \ref{contextual}. b); (2) many devices and services, within the subspace, are likely to be able to also influence the physical property. The choice of the most pertinent devices and services, given the purposes of the system, is not a trivial task (figure \ref{contextual}.c). In this context, maintaining the adequacy of the infrastructure with the purposes of ambient systems lies in their ability at reorganizing themselves (\kwrd{self-organized}) as the infrastructure evolves. From this point of view, middleware (\kwrd{Middleware}) are intended to organize the services involved and their relationships. They ensures the \textit{continuity of service}.

\input{contextual.tex}

\subsubsection{Technological factors}
\noindent Middleware implementation is only possible within an ecosystem of \textit{interoperable} services, both in terms of their accessibility within the network and in terms of the formal interpretation of the semantics of the interactions they offer with the physical environment. In this context, many challenges remain to be addressed.\\
\noindent The interoperability of the devices embedded on objects is primarily limited by the heterogeneity of the many communication protocols as no standard has clearly emerged to date. Hence, players in the field are grouped into consortiums around communication protocols, the main ones being Wi-Fi, Cellular (\kwrd{2G}, \kwrd{3G}, \kwrd{4G} and \kwrd{5G}), \kwrd{Bluetooth LE}, \kwrd{ZigBee}, \kwrd{KNX}, \kwrd{Z-Wave}, \kwrd{LoRaWAN}, \kwrd{enOcean}, etc.\\
\noindent On top of the communication protocols, application protocols based on service-oriented architectures (\kwrd{SOAP WS-*}), resources (\kwrd{REST API}, \kwrd{CoAp}) or messages (\kwrd{MQTT}, \kwrd{AMQP},...) are widely used. These application protocols have the advantage of providing a generic way to access the data and APIs of the devices through gateways and thus abstract the underlying technological heterogeneity. They also allow these devices to be integrated into the world wide web. Therefore, it is possible to use existing web technologies on devices and their data (i.e., search engines, indexing robots, semantic annotations and knowledge bases, security, composite applications, physical mashups, etc.). This is referred to as the web of objects (Web of Things, WoT). The formal description of the data and APIs of each of the devices can then be considered on the basis of semantic annotations (Semantic Web of Things, SWoT). With these annotations, middleware are then able to index and search for devices and their services in the environment, interpret their interactions and compose high added-value applications that implement multiple devices in a relevant way.\\
\noindent Originally syntactic, description formalisms such as \kwrd{RSDL} (RESTful Service Description Language), \kwrd{WSDL} (Web Service Description Language), etc. have evolved into semantic formalisms such as \kwrd{OWL-S} (Semantic Markup for Web Services), \kwrd{SAWSDL} (Semantic Annotations for WSDL), \kwrd{WSMO} (Web Service Modeling Ontology), etc. In general, semantic description formalisms are based on a vocabulary (an ontology). Interoperability is then only possible if all annotations are based on a common ontology. Unfortunately, it is not possible to describe semantically, \textit{a priori}, all objects, their devices, data and APIs within a single and global ontology (semantic heterogeneity) \cite{rocher2015run}. Hence, there are now a multitude of heterogeneous ontologies covering all fields of application. Thus, to date, 700 ontologies\footnote{\url{http://lov.okfn.org/dataset/lov/}}, which follow the W3C's recommendations in terms of quality and whose sources are available, are listed to date. This list does not take into account proprietary ontologies, many manufacturers of connected objects are likely to develop their own ontologies to describe objects, their devices, data and API.

\subsubsection{Environmental factors}
\noindent The nomadic nature of connected objects suggests that the vast majority of them cannot rely on wired power supply. Therefore, energy harvesting and storage strategies are implemented (\kwrd{self-powered}) \cite{murdan2018autonomous}\cite{ferrero2017multi}\cite{yue2017development}. In this context, at the heart of the concepts \kwrd{Zero Energy} and \kwrd{Green Computing}, wireless power transfer technologies (e.g., RFID tags) and solar energy recovery technologies are the most widely used. However, these technologies are largely dependent on the physical environment in which they are implemented. Thus, the performance of wireless energy transfer is highly dependent on the obstacles that can be found between the transmitter and receiver. The efficiency of solar panels, beyond technological considerations specific to the materials used to produce them, is largely dependent on weather variations and the condition of photovoltaic panels which, when exposed to dust, require frequent cleaning \cite{chaichan2018energy}. In addition, when batteries are used to store energy, they must be replaced regularly, as their lifespan is limited. Finally, some approaches use the piezoelectric effect to allow objects to produce their own energy \cite{babar2018energy}. This is the case, for example, with enOcean's switches, whose energy is created by the pressure exerted by the user. In terms of communication, the nomadic nature of connected objects also suggests the implementation of wireless communication technologies, again, largely dependent on the physical environment in which they are implemented. For example, temperature is a factor in the degradation of data transmission performance \cite{boano2010impact}\cite{marfievici2013environmental}.

\subsubsection{Human factors}
\noindent The vulnerability of ambient systems to cyber attacks (\kwrd{Cyber Security}, \kwrd{Cyber Attacks}, \kwrd{Crime}, \kwrd{Network Security}) could also compromise the integrity of these systems \cite{al2018cyber}. Whether by accidental or malicious modification of sensors data (e.g., man-in-the-middle attack), by failure of a device (e.g. Denial of Service attack, DoS, malfunction, etc.) or by its malicious takeover, the purposes of an ambient system can be compromised or no longer be achieved, with potential risks for individuals and their environment \cite{dudek2019cyber}.
\smallskip

\noindent Before closing the aspects related to the complexity of ambient systems, it is necessary to study the processes at the heart of the genesis of their decisions and which contribute to their complexity.

\subsection{The challenge of the Simonian rationality}
\noindent When ambient systems acquire a certain level of autonomy, their actions and the resulting effects are likely to be the result of bilateral or unilateral decisions \textbf{which result from algorithmic reasoning} (\kwrd{Decision Support Systems}, \kwrd{Decision Making}). Hence, the genesis of the decisions made by ambient systems and their rationality determine the achievement of the purposes of these systems. In this context, Herbert Simon's work, on the basis of Thomas Kuhn's work on Philosophy of Science, is particularly relevant. Economist (Nobel Prize in Economics in 1978) and sociologist, his work on limited rationality led him to focus on decision-making organizations and procedures as well as artificial intelligence (in 1975, with Allen Newell, he was awarded the Turing Prize for his work in this field). It is within the framework of this work that Herbert Simon introduces the concepts of \textit{substantive rationality} and \textit{procedural rationality} \cite{simon1976substantive}\cite{le1993capacite}.

\subsubsection{\textbf{Substantive rationality}}
\noindent Substantive (or optimizing) rationality refers to the behaviour of a system that makes its decisions while trying to optimize (\kwrd{Optimization}) the resources at its disposal in order to achieve its purposes. It deals with the adequacy between the means possessed by the agent and the purposes. \cite{simon2004sciences}. The rationality of decisions here is entirely determined by the characteristics of the environment in which they take place, i.e., the context. For instance, middleware demonstrate substantive rationality by selecting, from among all the available devices, those whose services are in adequacy with the purposes of the system. The selection process then follows an optimising logic, one of the criteria of which is the \textit{quality of service} (QoS, Quality of Service).

\subsubsection{\textbf{Procedural rationality}}
\noindent Procedural rationality is a decision-making process in which a system makes decisions according to predefined procedures. In this context, the rationality of the decisions taken is entirely determined by their algorithmic procedures, the improvement of which is necessarily accompanied by experimentation phases. Thus, the system examines, through a succession of trials and errors, a limited number of possible decisions and decides on the one that allows it to reach not the optimum but a minimum level of satisfaction. Procedural rationality highlights three essential phenomena that may call into question the achievement of the purposes of ambient systems:
\begin{enumerate}
  \setlength\itemsep{0.1em}
  \item Due to their limited computational capabilities and the complexity of the physical environment, ambient systems cannot claim to be optimal in their decisions,
  \item The set of choices facing ambient systems is finite, and, based on \cite{amendola1988dynamique}, \say{\textit{"all the choices available to the systems are not exogenous but endogenous to their past activity}}. As a result, the processes of learning (\kwrd{Learning Systems}) and decision-making are mutually constrained and difficult to distinguish,
  \item Ambient systems are likely to be highly heterogeneous, not only in terms of the means they use to achieve their goals, but also in terms of their operating environment and the local nature of their learning.
\end{enumerate}
\medskip

\noindent The arguments developed in this paragraph illustrate the intrinsically complex nature of ambient systems. This complexity is embodied in the impossibility of predicting their behaviour and therefore the achievement of their purposes, with the risks that this entails for both humans and their environment. At best, can we hope to get close to the desired behaviour but never reach it exactly: \say{\textit{[...] as soon as a system is open there is no optimum and any equilibrium is in interaction with its environment}} \cite{rochet:tel-00202800}. As a result, responsibility for actions is often left to individuals, with ambient systems only providing support for decision-making.

\section{Towards an epistemological rupture}
\label{epistemic}
\noindent The situation described in the previous paragraphs calls for an epistemological rupture that some researchers have advocated for \cite{bennaceur2019modelling}\cite{celik2019verifying}\cite{garlan2010software}\cite{lee2015past}\cite{bures2015software}. Indeed, current software engineering approaches, which include formal testing and verification methodologies (\kwrd{Formal Verification}), find their epistemological foundations in \textit{positivism}, a doctrine theorized by Auguste Comte in the continuity of Cartesianism, which claims for predictability. But, as indicated by Jean-Louis Le Moigne,  \say{\textit{In practice, analytical modelling is increasingly proving inadequate, whenever it is agreed that one is not sure that something cannot be forgotten (the hypothesis of closing the model), that objective evidence is only evident in a given ideology (...), in other words, whenever one has to make the assumption that the modelled phenomenon is not complicated but complex}} \cite{le1990modelisation} p.19.

\input{systemic.tex}

\noindent The predictability is assumed in the field of software engineering and justified by the technological means implemented to control computer systems and maintain their organisation in permanent adequacy with their purposes. This is not the case with ambient systems for which the notions of dependability \cite{avizienis2001fundamental} and trustworthiness, condition the acceptability of these systems.
\smallskip

\noindent Pragmatically, without predictability and being only able to get close to the desired behaviour without reaching it exactly, a relevant question to ask is then \say{\textit{to what extent does the ambient system actually do what it must do ?}}, the answer to which can only be given \textit{in vivo}. In order to address this question, it may be interesting to consider the \textit{systemic approach} . This approach finds its epistemological foundations in \textit{constructivism} \cite{le1995epistemologies}. It is then a question of considering an ambient system as a whole by focusing on the purposes it must produce (teleological point of view) in interaction with its environment (phenomenological point of view) without \textit{a priori} knowledge of its constituents.
\medskip

\noindent Systemic modelling consists in \say{\textit{representing a complex phenomenon as and by a system, a system that is general and stable enough to account for all the types of complexity that can be considered}} \cite{le1990modelisation} p.38. Although it is not possible to reduce complex systems to explanatory finite models, they still remain intelligible. Unlike analytical modelling, associated with disjunctive logic, systemic modelling is associated with conjunctive logic, which makes it possible to account for the perceived phenomena in and through their complex conjunctions (homomorphic correspondence between the object and its model, and not isomorphic as is the case with analytical modelling approaches). This translates into considering a set of actions characterized by the general notion of \textit{active processes} acting on the environment and characterized by inputs and outputs. On these basis, the figure \ref{systemic} depicts the evolution of a complex system in the (T)ransform-(S)pace-(F)orm referential.
\medskip

\noindent The systemic approach is a modelling method that takes place in iterative steps \cite{donnadieu2003approche}. The objective of these different steps is to obtain a consensual understanding of the complex system considered:
\begin{enumerate}
  \setlength\itemsep{0.1em}
  \item The first step consists in identifying the complex system to be studied ("the observable"), isolate it to define its contours (systemic breakdown) according to different criteria that remain at the discretion of the modeler,
  \item Using the \textit{triangulation method} \cite{le1994theorie} p.64, the complex system thus defined is then approached from the functional, structural and historical points of view. The functional aspect consists in questioning the purposes of the system (\say{\textit{What does it do?}}). The structural aspect focuses on studying the elements that make up the system by focusing on their interactions rather than on the elements themselves. Finally, the historical aspect consists in taking an interest in the temporal evolutions of the system. The knowledge of the system will then be represented by a barycentre whose position can evolve throughout the process of apprehending the object and improving its knowledge,
  \item This knowledge, once ordered, makes it possible to build a \textit{qualitative model} of the complex system (figure \ref{example}.a),
  \item Once the qualitative model has been constructed, it must be formally represented by qualifying the forms of inputs and outputs (non-functional properties) that are involved in the evolution of the system in a \textit{quantitative} manner \ref{example}.b). This formalization then allows the implementation of the model through temporal simulations. 
\end{enumerate}
\smallskip

\noindent By successive iterations, the model can be modified and improved until it reaches a consensus. A simple example is depicted in figure \ref{example}.
\medskip

\begin{figure}[ht!]
  \centering
  \begin{subfigure}[b]{0.5\textwidth}
      \centering
      \begin{tikzpicture}[->,>=stealth',shorten >=1pt,node distance=3.5cm,on grid,auto,thick,scale=0.75, every node/.style={scale=0.75},state/.style={circle, draw, minimum size=2.3cm}]
        \node[state,align=center] (LL) {$\overline{Lighted}$};
        \node[state,align=center] (HL) [right=of LL] {$Lighted$};
        \path[->] (LL) edge [loop above] node[align=center] {$\overline{presence}$} (LL);
        \path[->] (HL) edge [loop above] node[align=center] {$presence$} (HL);
        \path[->, bend left=20] (LL) edge node[align=center] {$presence$} (HL);
        \path[->, bend left=20] (HL) edge node[align=center] {$\overline{presence}$} (LL);	
      \end{tikzpicture}	
      \caption{\label{qualitatif} Qualitative model.}
  \end{subfigure}

  \begin{subfigure}[b]{0.5\textwidth}
      \centering
      \begin{tikzpicture}[->,>=stealth',shorten >=1pt,node distance=4.5cm,on grid,auto,thick,scale=0.75, every node/.style={scale=0.75},state/.style={circle, draw, minimum size=2.3cm}]
        \node[state, label=left:$\overline{Lighted}$,align=center] (LL) {$lum < 5.0$\\(output)};
        \node[state, label=right:$Lighted$,align=center] (HL) [right=of LL] {$lum > 25.0$\\(output)};
        \path[->] (LL) edge [loop above] node[align=center] {$pres < 3.0$ (input)} (LL);
        \path[->] (HL) edge [loop above] node[align=center] {$pres > 20.0$ (input)} (HL);
        \path[->, bend left=20] (LL) edge node[label=below:$duration \leq 5$,align=center] {$pres > 20.0$(input)} (HL);
        \path[->, bend left=20] (HL) edge node[label=above:$duration \leq 7$,align=center] {$pres < 3.0$(input)} (LL);	
      \end{tikzpicture}	
      \caption{\label{quantitatif} Quantitative model.}
  \end{subfigure}
  \caption{\label{example} Systemic modelling of a lighting control system. }
\end{figure}

\noindent Based on this model and the observation of inputs and outputs, it will therefore be necessary to assess the \textit{effectiveness} of an ambient system, i.e. the degree of adequacy between its purposes and the effects it produces. The effectiveness assessment is part of the Gibert's performance model as depicted in figure \ref{gibert}.
Per Jean-Louis Le Moigne, \say{"\textit{effectiveness is assessed by a multidimensional vector relating the behaviour of a system to its purposes}} \cite{le1990modelisation}.

\input{gibert.tex}

\noindent It is worth noting that middlewares act mainly at the relevance level of the Gibert model. For example, it is a matter of choosing the most relevant services (means) from among those available on the basis of their semantic annotations. Self-adaptive systems, on the other hand, generally act on low-level performance indicators that are part of an optimization problem at the efficiency level of Gibert's model (e.g. self-healing, self-configuring, self-optimizing, self-tuning, etc.). The same applies to evaluations relating to the concepts of dependability, performance, performability and quality of service (QoS) that focus mainly on low level objective properties, which need to be optimized over time.
\smallskip

\noindent However, \say{\textit{the efficiency of a vehicle can be measured by the ratio: 5 litres per 100 km; 5 litres of gasoline, resources consumed; 100 km travelled, resources produced. But if the purpose of the vehicle was to go to Paris and it ends up in Bordeaux, the measurement of its efficiency (a priori very good) will not say anything about the quality of its effectiveness.}} \cite{le1990modelisation}.     

\section{Perspectives}
\label{conclusion}

\noindent The evaluation of effectiveness, in addition to formal testing and verification methods, may offer interesting perspectives. Some of them are described below.

\subsection{Self-Adaptation}
\noindent Ambient systems are often self-adaptive, they are able, through feedback loops, to modify themselves in response to changes in their environment or the objectives they must achieve \cite{kephart2003vision} (e.g. Model-Analyze-Plan-Execute over a shared Knowledge (MAPE-K) \cite{kephart2003vision}, Model Identification MIAC and Model Reference Adaptive Controllers (MRAC) \cite{brun2009engineering}). In practice however, existing self-adaptive ambient systems are mainly designed by focusing on low-level performance indicators rather than high-level requirements \cite{peng2010self}\cite{macias2013self} (e.g., \textit{self-healing}, \textit{self-configuring}, \textit{self-optimizing}, \textit{self-tuning}, etc.).\\
\noindent Other approaches exist, based on reference models whose parameters are adjusted over time and on which formal verification methodologies are applied (Model Identification Adaptive Control (MIAC) and Model Reference Adaptive Control (MRAC)\cite{brun2009engineering}). These approaches, however, are only relevant if the models are in line with reality (i.e. are isomorph). 
\smallskip

\noindent \textbf{Equipped with continuous evaluation of their effectiveness, these systems would be able to focus on a high-level performance indicator that does not assume the illusory existence of an isomorphic model}.

\subsection{DevOps}

\noindent The DevOps approach \cite{sharma2017devops} seeks to unify the practices of software development (i.e., development, integration and testing) and systems administration actors (i.e., deployment, operation and maintenance). This unification is justified by the antagonistic objectives of these actors. On the one hand, software development actors are constrained by cost and time constraints with the negative impacts that this can have on the quality of the software delivered. On the other hand, IT administration actors are constrained to stability and quality objectives even if this has repercussions in terms of costs and delays.
\noindent The effectiveness assessment is part of a set of recommendations for the implementation of the \textsc{DevOps} approach. In particular \cite{sharma2017devops}:
\begin{enumerate}
  \setlength\itemsep{0.0em}
  \item \textbf{Continuous integration including continuous testing}. The idea of using the evaluation of effectiveness as a result of unit tests (which do not sanction execution with a PASS/FAIL binary result but with a degree of compliance) is fully justified here. They contribute to the dichotomy between positivist test approaches based on models of the systems considered and implemented at development time (\textsc{Dev}/Tests) and constructivist approaches based on observations (\textsc{Ops}, monitoring),
  \item \textbf{Users's feedback}. The qualitative model as depicted in figure \ref{example} may denote \textbf{users' preferences}. In this context, the evaluation of the effectiveness could be used as an automatic users' feedback indicator,
  \item \textbf{Monitoring of operation and production quality using key metrics and indicators}. As such, the effectiveness assessment could provide such indicators. 
\end{enumerate}	    

\subsection{Quality of Experience}

\noindent The Quality of Experience (QoE) provides an assessment of end-users satisfaction with a service or, more generally, a product \cite{crespi2011qoe}. It aims to take into account all factors that contribute to the quality of a system or service as perceived by the users, beyond purely technological factors. A definition is provided by the International Telecommunication Union (ITU) : \say{\textit{The degree of delight or annoyance of the user of an application or service. It results from the fulfillment of his or her expectations with respect to the utility and/or enjoyment of the application or service in the light of the user’s personality and current state.}}
\smallskip

\noindent In order to measure the quality of experience, human evaluations can be used. In this context, the Mean Opinion Score (MOS \cite{streijl2016mean}) is widely used. Évaluations are mainly governed by ITU recommendations which define strict experimental frameworks for multimedia application areas (communication, video, speech, etc.). Besides MOS, some approaches are based on Random Neural Networks (RNN) as, for instance, the Pseudo Subjective Quality Assessment (PSQA) \cite{rubino2006quantifying} which involves learning, through a set of subjective tests, how humans respond to quality.
\smallskip

\noindent MOS is only a limited form of quality of experience assessment, acquired in controlled environments that cannot reflect the open nature of environments where users' experiences are shaped \cite{wechsung2014quality}. RNN-based approaches require a significant number of annotated datasets with the estimation of their respective quality. In the context of ambient systems, datasets must be acquired in different real situations, the number of which is likely to be prohibitive. In addition, given the subjective nature of the QoE assessment, this number should be multiplied by an equally large number of users. This approach therefore seems impracticable when it comes to generalizing it to ambient systems.
\smallskip

\noindent \textbf{Here again, assuming that the qualitative model as depicted in figure \ref{example} may denote users' preferences, the evaluation of the effectiveness could be used as a QoE indicator}.

\subsection{Dependability}

\noindent The notion of dependability \cite{avizienis2001fundamental} covers all critical aspects related to the reliability of ambient systems \cite{eusgeld2008dependability}. A definition of dependability, given by the International Federation for Information Processing (IFIP), is the following:\say{\textit{The trustworthiness of a computing system which allows reliance to be justifiably placed on the service it delivers, enables these various concerns to be subsumed within a single conceptual framework. Dependability thus includes as special cases such attributes as reliability, availability, safety, security.}}
\smallskip

\noindent  IEEE defines \textit{reliability} as the \say{\textit{ability of a system or component to perform its required functions under stated conditions for a specified period of time}} (continuity of correct service). \textit{Availability} is concerned with the ability of a system to provide a service when it is required. \textit{Safety} is about the ability of a system to operate normally or abnormally, without adverse effects on humans or the environment. Finally, \textit{security} is the ability of a system to prevent unauthorized access or manipulation of its state.
\smallskip

\noindent The notion of dependability is often associated with the notions of \textit{performance} and \textit{performability}. IEEE defines the performance as \say{\textit{the degree to which a system or component accomplishes its designated functions within given constraints, such as speed, accuracy or memory usage.}}. The performability \say{\textit{quantifies how well the object system performs in the presence of faults over a specified period of time}}\cite{meyer1993specification}.
\medskip

\noindent When these attributes are not based on non-functional indicators, they require an implicit assessment of the behaviour of the system under consideration. This includes, for instance, measuring Mean Down Time (MDT), Mean Time To Failure (MTTF), Mean Time Before Failure (MTBF), etc. \textbf{In this context, the evaluation of effectiveness can be used as a high-level indicator for both dependability and performance analysis, beyond non-functional indicators}.

\section{Conclusion}

\noindent Their purposes being realized in the physical environment, ambient systems are complex systems that are embodied in a set of concepts such as Pervasive Computing, Ubiquitous Computing, Cyber-Physical Systems, Ambient Intelligence, etc. This complexity, by resulting in the inability to obtain a predictive isomorphic model of the effects of their actions, implies that the responsibility for them often lies with users. When this is not the case, the effects produced are not without risk for users and their environment. 
In this context, we have proposed the systemic modelling approach, which has its epistemological roots in constructivism: it is no longer a question of predicting the effects produced by these systems but, in a more pragmatic way, of evaluating their effectiveness on the basis of a model of their purposes.
\medskip

\noindent The proposed modelling approach paves the way to a set of opportunities in the areas of auto-adaptive systems, user experience, dependability, etc. First results are encouraging. For instance, the proposed modelling approach has been implemented in \cite{rocher2017probabilistic} where effectiveness assessment is achieved using probabilistic Input/Output Hidden Markov Model (IOHMM) and extended into the possibilistic framework in \cite{rocher2018possibilistic}, accounting for temporal constraints. 

\section*{Acknowledgment}
\noindent This work has been conjointly supported by GFI Informatique, Innovation group, Saint-Ouen, France and the French National Center for Scientific Research (CNRS), Sophia Antipolis, France.

\bibliographystyle{IEEEtran}
\bibliography{references}

\end{document}

%% file: applications.tex
\definecolor{g}{HTML}{3366ff}
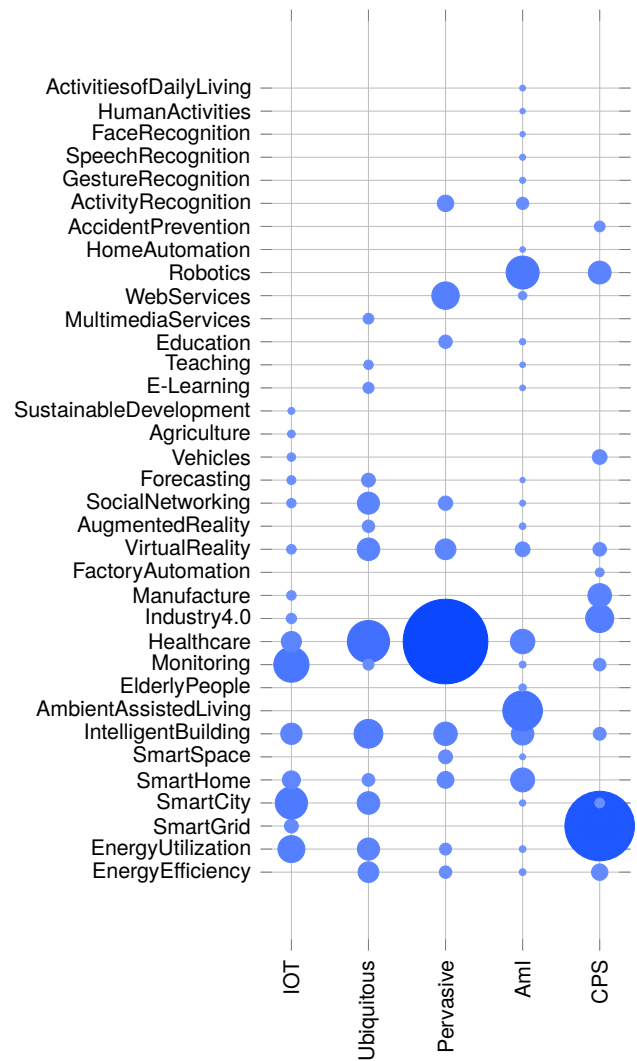
\begin{figure}[!ht]
\centering
\pgfplotsset{
  tick label style = {font=\sffamily\small},
  every axis label = {font=\sffamily\small},
  legend style = {font=\sffamily\small},
  label style = {font=\sffamily\small}
}
    \begin{tikzpicture}   
        \begin{axis}[  
                enlargelimits=true,
                width=6.5cm,
                height=14.1cm,        
                xticklabel style={rotate=90},
                yticklabel style={rotate=0},
                grid=major,
                xtick=data,
                ytick=data,
                point meta=explicit,
                axis line style={draw=none},
                scatter/@pre marker code/.code={
                    \pgfmathparse{                                  
                        \pgfplotspointmetatransformed/1000*50+70}   
                    \let\opacity=\pgfmathresult                     
                    \pgfmathparse{                                  
                        \pgfplotspointmetatransformed/1000*15+1}   
                    \def\markopts{                                  
                        mark=*,                                     
                        color=g!\opacity,                     
                        fill=g!\opacity,                      
                        mark size=\pgfmathresult}                   
                    \expandafter\scope\expandafter[\markopts]},
                scatter/@post marker code/.code={
                    \endscope},
                symbolic y coords={
EnergyEfficiency,
EnergyUtilization,
SmartGrid,
SmartCity,
SmartHome,
SmartSpace,
IntelligentBuilding,
AmbientAssistedLiving,
ElderlyPeople,
Monitoring,
Healthcare,
Industry4.0,
Manufacture,
FactoryAutomation,
VirtualReality,
AugmentedReality,
SocialNetworking,
Forecasting,
Vehicles,
Agriculture,
SustainableDevelopment,
E-Learning,
Teaching,
Education,
MultimediaServices,
WebServices,
Robotics,
HomeAutomation,
AccidentPrevention,
ActivityRecognition,
GestureRecognition,
SpeechRecognition,
FaceRecognition,
HumanActivities,
ActivitiesofDailyLiving,                  
                            },
                symbolic x coords={
                                    IOT,
                                    Ubiquitous,
                                    Pervasive,
                                    AmI,
                                    CPS
                            },
            ]
            \addplot[only marks,scatter]
                table[x index=1, y index=0, meta index=2] {Apps2019.csv};
        \end{axis}
    \end{tikzpicture}   
    \caption{\label{apps}Application domains associated with the concepts (each row sums to 100\%).}
\end{figure}

%% file: chronologie.tex
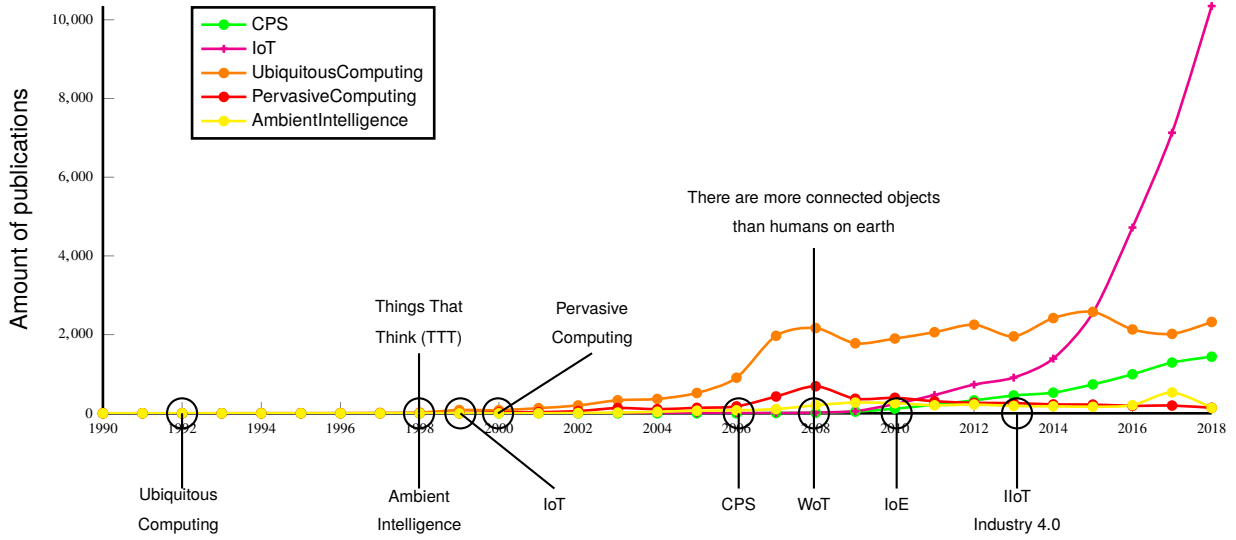
\begin{figure*}[ht]
\centering
\pgfplotsset{every tick label/.append style={font=\tiny}}
\begin{tikzpicture}[thick,scale=1.0, every node/.style={scale=1.0},font=\sffamily]
  \begin{axis}
    [
     xticklabel style={/pgf/number format/.cd,%
      scaled x ticks = false, 
      set thousands separator={}, 
      fixed},%
      scaled y ticks=false,
     enlargelimits=false,
     legend style={nodes={scale=0.75, transform shape}},
     width=0.9\linewidth,
     height=7cm,
     grid=none,
     axis lines*=left,
     ylabel=Amount of publications,
     legend cell align={left},
     line width=1.pt,
     mark size=1.5pt,
     legend style={at={(0.3,1)},anchor=north east,draw=black,fill=white,align=left},
    ]

    \addplot[color=green,solid,mark=*,mark options={solid},smooth]
    table [x=Year,y=CPS]{years.csv};
    
    \addplot[color=magenta,solid,mark=+,mark options={solid},smooth] 
    table [x=Year,y=IoT]{years.csv};
    
    \addplot[color=orange,solid,mark=*,mark options={solid},smooth]
    table [x=Year,y=UbiquitousComputing]{years.csv};

    \addplot[color=red,solid,mark=*,mark options={solid},smooth] 
    table [x=Year,y=PervasiveComputing]{years.csv};
    
    \addplot[color=yellow,solid,mark=*,mark options={solid},smooth] 
    table [x=Year,y=AmbientIntelligence]{years.csv};
    
    
    
    \legend{CPS,IoT,UbiquitousComputing,PervasiveComputing,AmbientIntelligence}
  \end{axis}
  
   \node at (1,-1.5)    {\scriptsize Computing}; \node at (1,-1.1) {\scriptsize Ubiquitous};
   \node at (4.2,-1.5)  {\scriptsize Intelligence};\node at (4.2,-1.1) {\scriptsize Ambient};
   \node at (6,-1.2)    {\scriptsize IoT};
   \node at (8.45,-1.2) {\scriptsize CPS};
   \node at (10.55,-1.2) {\scriptsize IoE};
   \node at (4.2,1)     {\scriptsize Think (TTT)}; \node at (4.2,1.4) {\scriptsize Things That};
   \node at (6.5,1)     {\scriptsize Computing}; \node at (6.5,1.4) {\scriptsize Pervasive};

   \node at (9.45,2.5) {\scriptsize than humans on earth}; \node at (9.45,2.85) {\scriptsize There are more connected objects};
   \node at (9.45,-1.2)  {\scriptsize WoT};

   \node at (12.15,-1.1)  {\scriptsize IIoT};
   \node at (12.15,-1.5)  {\scriptsize Industry 4.0};

   \draw (1.05,0)  -- (1.05,-1);        
   \draw (4.2,0)   -- (4.2,-1);      
   \draw (4.2,0)   -- (4.2,.8);    
   \draw (4.75,0)  -- (6,-1);        
   \draw (5.25,0)  -- (6.5,.8);      
   \draw (8.45,0)  -- (8.45,-1);     
   \draw (9.45,0)   -- (9.45,-1);      
   \draw (9.45,0)   -- (9.45,2.2);   
   \draw (10.55,0)  -- (10.55,-1);     
   \draw (12.15,0) -- (12.15,-1);    

   \draw (1.05,0)     circle (0.2cm);   
   \draw (4.2,0)   circle (0.2cm);   
   \draw (4.75,0)  circle (0.2cm);   
   \draw (5.25,0)  circle (0.2cm);   
   \draw (8.45,0)  circle (0.2cm);   
   \draw (9.45,0)   circle (0.2cm);   
   \draw (10.55,0)  circle (0.2cm);   
   \draw (12.15,0) circle (0.2cm);   

 \end{tikzpicture}  
 \caption{\label{evolution} \small Chronology of appearance of terms relating to ambient systems.}
\end{figure*}

%% file: object.tex
\begin{figure*}[!ht]
    \centering
    
    \tikzset{every picture/.style={line width=0.75pt}} 
    \begin{tikzpicture}[thick,scale=0.65, every node/.style={scale=0.8}, x=0.75pt,y=0.75pt,yscale=-1,xscale=1,font=\sffamily]

    \draw  [line width=2.25]  (566.05,267.4) .. controls (566.05,261.66) and (570.71,257) .. (576.45,257) -- (719.8,257) .. controls (725.54,257) and (730.2,261.66) .. (730.2,267.4) -- (730.2,298.6) .. controls (730.2,304.34) and (725.54,309) .. (719.8,309) -- (576.45,309) .. controls (570.71,309) and (566.05,304.34) .. (566.05,298.6) -- cycle ;
    
    \draw   (244,443.4) .. controls (244,437.66) and (248.66,433) .. (254.4,433) -- (349.1,433) .. controls (354.84,433) and (359.5,437.66) .. (359.5,443.4) -- (359.5,474.6) .. controls (359.5,480.34) and (354.84,485) .. (349.1,485) -- (254.4,485) .. controls (248.66,485) and (244,480.34) .. (244,474.6) -- cycle ;
    \draw   (368.8,443.2) .. controls (368.8,437.46) and (373.46,432.8) .. (379.2,432.8) -- (473.9,432.8) .. controls (479.64,432.8) and (484.3,437.46) .. (484.3,443.2) -- (484.3,474.4) .. controls (484.3,480.14) and (479.64,484.8) .. (473.9,484.8) -- (379.2,484.8) .. controls (373.46,484.8) and (368.8,480.14) .. (368.8,474.4) -- cycle ;
    \draw   (243.6,559) .. controls (243.6,553.26) and (248.26,548.6) .. (254,548.6) -- (348.7,548.6) .. controls (354.44,548.6) and (359.1,553.26) .. (359.1,559) -- (359.1,590.2) .. controls (359.1,595.94) and (354.44,600.6) .. (348.7,600.6) -- (254,600.6) .. controls (248.26,600.6) and (243.6,595.94) .. (243.6,590.2) -- cycle ;
    
    \draw   (570.6,559.2) .. controls (570.6,553.46) and (575.26,548.8) .. (581,548.8) -- (675.7,548.8) .. controls (681.44,548.8) and (686.1,553.46) .. (686.1,559.2) -- (686.1,590.4) .. controls (686.1,596.14) and (681.44,600.8) .. (675.7,600.8) -- (581,600.8) .. controls (575.26,600.8) and (570.6,596.14) .. (570.6,590.4) -- cycle ;
    
    \draw   (804,440.4) .. controls (804,434.66) and (808.66,430) .. (814.4,430) -- (909.1,430) .. controls (914.84,430) and (919.5,434.66) .. (919.5,440.4) -- (919.5,471.6) .. controls (919.5,477.34) and (914.84,482) .. (909.1,482) -- (814.4,482) .. controls (808.66,482) and (804,477.34) .. (804,471.6) -- cycle ;
    
    \draw   (308,711.6) .. controls (308,705.86) and (312.66,701.2) .. (318.4,701.2) -- (413.1,701.2) .. controls (418.84,701.2) and (423.5,705.86) .. (423.5,711.6) -- (423.5,742.8) .. controls (423.5,748.54) and (418.84,753.2) .. (413.1,753.2) -- (318.4,753.2) .. controls (312.66,753.2) and (308,748.54) .. (308,742.8) -- cycle ;
    
    \draw   (307.2,802.4) .. controls (307.2,796.66) and (311.86,792) .. (317.6,792) -- (412.3,792) .. controls (418.04,792) and (422.7,796.66) .. (422.7,802.4) -- (422.7,833.6) .. controls (422.7,839.34) and (418.04,844) .. (412.3,844) -- (317.6,844) .. controls (311.86,844) and (307.2,839.34) .. (307.2,833.6) -- cycle ;
    
    \draw    (426.5,299) -- (426.5,433) ;

    \draw [color={rgb, 255:red, 0; green, 0; blue, 0 }  ,draw opacity=1 ][fill={rgb, 255:red, 255; green, 255; blue, 255 }  ,fill opacity=1 ]   (303.5,279) -- (304,433) ;

    \draw [color={rgb, 255:red, 0; green, 0; blue, 0 }  ,draw opacity=1 ][fill={rgb, 255:red, 255; green, 255; blue, 255 }  ,fill opacity=1 ]   (302.6,486) -- (302.7,549.2) ;
    
    \draw [shift={(302.6,484)}, rotate = 89.91] [color={rgb, 255:red, 0; green, 0; blue, 0 }  ,draw opacity=1 ][line width=0.75]    (10.93,-3.29) .. controls (6.95,-1.4) and (3.31,-0.3) .. (0,0) .. controls (3.31,0.3) and (6.95,1.4) .. (10.93,3.29)   ;
    \draw [color={rgb, 255:red, 0; green, 0; blue, 0 }  ,draw opacity=1 ][fill={rgb, 255:red, 255; green, 255; blue, 255 }  ,fill opacity=1 ]   (359.5,573.8) -- (423.43,574) ;

    \draw [line width=0.75]    (427.5,627) -- (427.5,485) ;

    \draw [line width=0.75]    (302.7,626.2) -- (302.7,600.2) ;

    \draw [line width=0.75]    (302.7,626.2) -- (942.5,626) ;

    \draw [line width=0.75]    (365.5,700) -- (365.6,627.1) ;
    
    \draw [shift={(365.5,702)}, rotate = 270.08] [color={rgb, 255:red, 0; green, 0; blue, 0 }  ][line width=0.75]    (10.93,-3.29) .. controls (6.95,-1.4) and (3.31,-0.3) .. (0,0) .. controls (3.31,0.3) and (6.95,1.4) .. (10.93,3.29)   ;
    \draw [color={rgb, 255:red, 0; green, 0; blue, 0 }  ,draw opacity=1 ][fill={rgb, 255:red, 255; green, 255; blue, 255 }  ,fill opacity=1 ]   (365.6,756.1) -- (365.6,792.1) ;
    
    \draw [shift={(365.6,754.1)}, rotate = 90] [color={rgb, 255:red, 0; green, 0; blue, 0 }  ,draw opacity=1 ][line width=0.75]    (10.93,-3.29) .. controls (6.95,-1.4) and (3.31,-0.3) .. (0,0) .. controls (3.31,0.3) and (6.95,1.4) .. (10.93,3.29)   ;
    \draw [color={rgb, 255:red, 0; green, 0; blue, 0 }  ,draw opacity=1 ][fill={rgb, 255:red, 255; green, 255; blue, 255 }  ,fill opacity=1 ]   (364.51,846) -- (364.8,904.4) ;
    
    \draw [shift={(364.5,844)}, rotate = 89.72] [color={rgb, 255:red, 0; green, 0; blue, 0 }  ,draw opacity=1 ][line width=0.75]    (10.93,-3.29) .. controls (6.95,-1.4) and (3.31,-0.3) .. (0,0) .. controls (3.31,0.3) and (6.95,1.4) .. (10.93,3.29)   ;
    \draw    (802.5,458) -- (484.5,458) ;
    
    \draw [shift={(804.5,458)}, rotate = 180] [color={rgb, 255:red, 0; green, 0; blue, 0 }  ][line width=0.75]    (10.93,-3.29) .. controls (6.95,-1.4) and (3.31,-0.3) .. (0,0) .. controls (3.31,0.3) and (6.95,1.4) .. (10.93,3.29)   ;
    \draw   (804.6,559.2) .. controls (804.6,553.46) and (809.26,548.8) .. (815,548.8) -- (909.7,548.8) .. controls (915.44,548.8) and (920.1,553.46) .. (920.1,559.2) -- (920.1,590.4) .. controls (920.1,596.14) and (915.44,600.8) .. (909.7,600.8) -- (815,600.8) .. controls (809.26,600.8) and (804.6,596.14) .. (804.6,590.4) -- cycle ;
    
    \draw [line width=0.75]    (861.7,627.2) -- (861.7,601.2) ;

    \draw [color={rgb, 255:red, 0; green, 0; blue, 0 }  ,draw opacity=1 ][fill={rgb, 255:red, 255; green, 255; blue, 255 }  ,fill opacity=1 ]   (881.5,484) -- (881.5,549) ;
    
    \draw [shift={(881.5,482)}, rotate = 90] [color={rgb, 255:red, 0; green, 0; blue, 0 }  ,draw opacity=1 ][line width=0.75]    (10.93,-3.29) .. controls (6.95,-1.4) and (3.31,-0.3) .. (0,0) .. controls (3.31,0.3) and (6.95,1.4) .. (10.93,3.29)   ;
    \draw [color={rgb, 255:red, 0; green, 0; blue, 0 }  ,draw opacity=1 ][fill={rgb, 255:red, 255; green, 255; blue, 255 }  ,fill opacity=1 ]   (838.6,483) -- (838.7,549.2) ;
    
    \draw [shift={(838.6,481)}, rotate = 89.92] [color={rgb, 255:red, 0; green, 0; blue, 0 }  ,draw opacity=1 ][line width=0.75]    (10.93,-3.29) .. controls (6.95,-1.4) and (3.31,-0.3) .. (0,0) .. controls (3.31,0.3) and (6.95,1.4) .. (10.93,3.29)   ;
    \draw [color={rgb, 255:red, 0; green, 0; blue, 0 }  ,draw opacity=1 ][fill={rgb, 255:red, 255; green, 255; blue, 255 }  ,fill opacity=1 ]   (860.6,389) -- (860.7,429.2) ;
    
    \draw [shift={(860.6,387)}, rotate = 89.86] [color={rgb, 255:red, 0; green, 0; blue, 0 }  ,draw opacity=1 ][line width=0.75]    (10.93,-3.29) .. controls (6.95,-1.4) and (3.31,-0.3) .. (0,0) .. controls (3.31,0.3) and (6.95,1.4) .. (10.93,3.29)   ;
    \draw   (802.6,347.2) .. controls (802.6,341.46) and (807.26,336.8) .. (813,336.8) -- (907.7,336.8) .. controls (913.44,336.8) and (918.1,341.46) .. (918.1,347.2) -- (918.1,378.4) .. controls (918.1,384.14) and (913.44,388.8) .. (907.7,388.8) -- (813,388.8) .. controls (807.26,388.8) and (802.6,384.14) .. (802.6,378.4) -- cycle ;
    
    \draw  [draw opacity=0] (422.84,574.63) .. controls (422.69,574.1) and (422.62,573.53) .. (422.63,572.94) .. controls (422.66,569.9) and (424.88,567.46) .. (427.57,567.5) .. controls (430.26,567.53) and (432.41,570.02) .. (432.37,573.06) .. controls (432.37,573.65) and (432.28,574.22) .. (432.12,574.75) -- (427.5,573) -- cycle ; \draw   (422.84,574.63) .. controls (422.69,574.1) and (422.62,573.53) .. (422.63,572.94) .. controls (422.66,569.9) and (424.88,567.46) .. (427.57,567.5) .. controls (430.26,567.53) and (432.41,570.02) .. (432.37,573.06) .. controls (432.37,573.65) and (432.28,574.22) .. (432.12,574.75) ;
    \draw [color={rgb, 255:red, 0; green, 0; blue, 0 }  ,draw opacity=1 ][fill={rgb, 255:red, 255; green, 255; blue, 255 }  ,fill opacity=1 ]   (432.12,574.75) -- (569.05,574.94) ;
    \draw [shift={(571.05,574.95)}, rotate = 180.08] [color={rgb, 255:red, 0; green, 0; blue, 0 }  ,draw opacity=1 ][line width=0.75]    (10.93,-3.29) .. controls (6.95,-1.4) and (3.31,-0.3) .. (0,0) .. controls (3.31,0.3) and (6.95,1.4) .. (10.93,3.29)   ;
    
    \draw [line width=0.75]    (942.6,360.2) -- (942.5,626) ;

    \draw [line width=0.75]    (942.8,359.8) -- (918.5,360) ;

    \draw  [fill={rgb, 255:red, 255; green, 255; blue, 255 }  ,fill opacity=1 ][line width=2.25]  (307,915.07) .. controls (307,909.32) and (311.66,904.67) .. (317.4,904.67) -- (412.1,904.67) .. controls (417.84,904.67) and (422.5,909.32) .. (422.5,915.07) -- (422.5,946.27) .. controls (422.5,952.01) and (417.84,956.67) .. (412.1,956.67) -- (317.4,956.67) .. controls (311.66,956.67) and (307,952.01) .. (307,946.27) -- cycle ;
    
    \draw  [dash pattern={on 4.5pt off 4.5pt}] (194.5,342.9) .. controls (194.5,334.67) and (201.17,328) .. (209.4,328) -- (943.6,328) .. controls (951.83,328) and (958.5,334.67) .. (958.5,342.9) -- (958.5,951.1) .. controls (958.5,959.33) and (951.83,966) .. (943.6,966) -- (209.4,966) .. controls (201.17,966) and (194.5,959.33) .. (194.5,951.1) -- cycle ;
    \draw    (304.5,354) -- (209.95,353.5) ;

    \draw [color={rgb, 255:red, 0; green, 0; blue, 0 }  ,draw opacity=1 ][fill={rgb, 255:red, 255; green, 255; blue, 255 }  ,fill opacity=1 ]   (560,279) -- (303.5,279) ;
    
    \draw [shift={(562,279)}, rotate = 180] [color={rgb, 255:red, 0; green, 0; blue, 0 }  ,draw opacity=1 ][line width=0.75]    (10.93,-3.29) .. controls (6.95,-1.4) and (3.31,-0.3) .. (0,0) .. controls (3.31,0.3) and (6.95,1.4) .. (10.93,3.29)   ;
    \draw   (572.2,802.4) .. controls (572.2,796.66) and (576.86,792) .. (582.6,792) -- (677.3,792) .. controls (683.04,792) and (687.7,796.66) .. (687.7,802.4) -- (687.7,833.6) .. controls (687.7,839.34) and (683.04,844) .. (677.3,844) -- (582.6,844) .. controls (576.86,844) and (572.2,839.34) .. (572.2,833.6) -- cycle ;
    
    \draw    (570.5,818) -- (422.5,818) ;
    
    \draw [shift={(572.5,818)}, rotate = 180] [color={rgb, 255:red, 0; green, 0; blue, 0 }  ][line width=0.75]    (10.93,-3.29) .. controls (6.95,-1.4) and (3.31,-0.3) .. (0,0) .. controls (3.31,0.3) and (6.95,1.4) .. (10.93,3.29)   ;
    \draw    (210.5,931) -- (306.5,931) ;

    \draw    (210.5,353) -- (210.5,931) ;

    \draw  [fill={rgb, 255:red, 0; green, 0; blue, 0 }  ,fill opacity=0.04 ][dash pattern={on 0.84pt off 2.51pt}] (235.6,400.05) .. controls (235.6,392.68) and (241.58,386.7) .. (248.95,386.7) -- (484.25,386.7) .. controls (491.62,386.7) and (497.6,392.68) .. (497.6,400.05) -- (497.6,598.65) .. controls (497.6,606.02) and (491.62,612) .. (484.25,612) -- (248.95,612) .. controls (241.58,612) and (235.6,606.02) .. (235.6,598.65) -- cycle ;
    \draw  [fill={rgb, 255:red, 0; green, 0; blue, 0 }  ,fill opacity=0.04 ][dash pattern={on 0.84pt off 2.51pt}] (769.27,343.67) .. controls (769.27,337.78) and (774.05,333) .. (779.94,333) -- (938.6,333) .. controls (944.49,333) and (949.27,337.78) .. (949.27,343.67) -- (949.27,636.33) .. controls (949.27,642.22) and (944.49,647) .. (938.6,647) -- (779.94,647) .. controls (774.05,647) and (769.27,642.22) .. (769.27,636.33) -- cycle ;
    \draw [color={rgb, 255:red, 0; green, 0; blue, 0 }  ,draw opacity=1 ][fill={rgb, 255:red, 255; green, 255; blue, 255 }  ,fill opacity=1 ]   (561,298.6) -- (426.5,298.6) ;
    
    \draw [shift={(563,298.6)}, rotate = 180] [color={rgb, 255:red, 0; green, 0; blue, 0 }  ,draw opacity=1 ][line width=0.75]    (10.93,-3.29) .. controls (6.95,-1.4) and (3.31,-0.3) .. (0,0) .. controls (3.31,0.3) and (6.95,1.4) .. (10.93,3.29)   ;
    \draw    (450.5,931) -- (422.5,931) ;

    \draw    (450.5,931) -- (450.5,877) ;

    \draw    (450.5,877) -- (401.5,877) ;

    \draw    (401.5,903) -- (401.5,877) ;
    
    \draw [shift={(401.5,905)}, rotate = 270] [color={rgb, 255:red, 0; green, 0; blue, 0 }  ][line width=0.75]    (10.93,-3.29) .. controls (6.95,-1.4) and (3.31,-0.3) .. (0,0) .. controls (3.31,0.3) and (6.95,1.4) .. (10.93,3.29)   ;
    \draw [color={rgb, 255:red, 0; green, 0; blue, 0 }  ,draw opacity=1 ][fill={rgb, 255:red, 255; green, 255; blue, 255 }  ,fill opacity=1 ]   (731.5,467) -- (802.43,467.19) ;
    \draw [shift={(804.43,467.2)}, rotate = 180.16] [color={rgb, 255:red, 0; green, 0; blue, 0 }  ,draw opacity=1 ][line width=0.75]    (10.93,-3.29) .. controls (6.95,-1.4) and (3.31,-0.3) .. (0,0) .. controls (3.31,0.3) and (6.95,1.4) .. (10.93,3.29)   ;
    
    \draw    (731.5,575) -- (686.5,575) ;

    \draw    (731.5,467) -- (731.5,575) ;

    \draw   (571.2,711.4) .. controls (571.2,705.66) and (575.86,701) .. (581.6,701) -- (676.3,701) .. controls (682.04,701) and (686.7,705.66) .. (686.7,711.4) -- (686.7,742.6) .. controls (686.7,748.34) and (682.04,753) .. (676.3,753) -- (581.6,753) .. controls (575.86,753) and (571.2,748.34) .. (571.2,742.6) -- cycle ;
    
    \draw   (804.2,710.4) .. controls (804.2,704.66) and (808.86,700) .. (814.6,700) -- (909.3,700) .. controls (915.04,700) and (919.7,704.66) .. (919.7,710.4) -- (919.7,741.6) .. controls (919.7,747.34) and (915.04,752) .. (909.3,752) -- (814.6,752) .. controls (808.86,752) and (804.2,747.34) .. (804.2,741.6) -- cycle ;
    
    \draw    (801.5,726) -- (686.5,726) ;
    
    \draw [shift={(803.5,726)}, rotate = 180] [color={rgb, 255:red, 0; green, 0; blue, 0 }  ][line width=0.75]    (10.93,-3.29) .. controls (6.95,-1.4) and (3.31,-0.3) .. (0,0) .. controls (3.31,0.3) and (6.95,1.4) .. (10.93,3.29)   ;
    \draw [color={rgb, 255:red, 0; green, 0; blue, 0 }  ,draw opacity=1 ][fill={rgb, 255:red, 255; green, 255; blue, 255 }  ,fill opacity=1 ]   (424.5,728) -- (571.05,727.95) ;
    
    \draw [shift={(422.5,728)}, rotate = 359.98] [color={rgb, 255:red, 0; green, 0; blue, 0 }  ,draw opacity=1 ][line width=0.75]    (10.93,-3.29) .. controls (6.95,-1.4) and (3.31,-0.3) .. (0,0) .. controls (3.31,0.3) and (6.95,1.4) .. (10.93,3.29)   ;
    \draw    (705.5,713) -- (705.5,672) ;

    \draw    (705.5,672) -- (656.5,672) ;

    \draw    (656.5,698) -- (656.5,672) ;
    
    \draw [shift={(656.5,700)}, rotate = 270] [color={rgb, 255:red, 0; green, 0; blue, 0 }  ][line width=0.75]    (10.93,-3.29) .. controls (6.95,-1.4) and (3.31,-0.3) .. (0,0) .. controls (3.31,0.3) and (6.95,1.4) .. (10.93,3.29)   ;
    \draw    (705.5,713) -- (685.5,713) ;

    \draw (426.55,458.8) node  [align=left] {Sensor};
    \draw (301.75,459) node  [align=left] {Effector};
    \draw (465.67,405.33) node  [align=left] {{\small observes}};
    \draw (265.27,406.13) node  [align=left] {modifies};
    \draw (542.87,270.93) node  [align=left] {1..*};
    \draw (311.27,423.53) node  [align=left] {1};
    \draw (420.27,424.13) node  [align=left] {1};
    \draw (341.27,515.53) node  [align=left] {acts on};
    \draw (293.47,537.93) node  [align=left] {1};
    \draw (286.27,495.33) node  [align=left] {1..*};
    \draw (556.87,564.53) node  [align=left] {1..*};
    \draw (366.07,564.53) node  [align=left] {1};
    \draw (351.07,661.13) node  [align=left] {est};
    \draw (373.67,783.13) node  [align=left] {1};
    \draw (373.47,761.53) node  [align=left] {*};
    \draw (410.47,772.13) node  [align=left] {contains};
    \draw (372.07,895.93) node  [align=left] {1};
    \draw (372.27,853.73) node  [align=left] {*};
    \draw (492.07,450.13) node  [align=left] {1};
    \draw (790.67,450.13) node  [align=left] {1..*};
    \draw (499.55,587.93) node  [align=left] {consumes};
    \draw (626.67,448.33) node  [align=left] {produces};
    \draw (713.07,517.13) node  [align=left] {is};
    \draw (889.67,540.13) node  [align=left] {1};
    \draw (900.87,490.53) node  [align=left] {1..*};
    \draw (912.67,514.33) node  [align=left] {produces};
    \draw (794.27,516.93) node  [align=left] {consumes};
    \draw (846.07,539.53) node  [align=left] {1};
    \draw (853.27,492.33) node  [align=left] {1..*};
    \draw (814.27,408.93) node  [align=left] {consumes};
    \draw (868.67,422.13) node  [align=left] {1};
    \draw (875.67,397.13) node  [align=left] {1..*};
    \draw (430.67,810.13) node  [align=left] {1};
    \draw (558.67,810.13) node  [align=left] {1..*};
    \draw (261.27,921.13) node  [align=left] {modifies};
    \draw (368.55,373.93) node  [align=left] {\textbf{interfaces}};
    \draw (861.55,657.93) node  [align=left] {\textbf{storage/treatment}};
    \draw (492.55,801.93) node  [align=left] {consumes};
    \draw (542.87,291.93) node  [align=left] {1..*};
    \draw (431.07,920.93) node  [align=left] {1};
    \draw (409.27,894.73) node  [align=left] {*};
    \draw (521.47,901.13) node  [align=left] {mechanically linked};
    \draw (501.55,713.93) node  [align=left] {consumes};
    \draw (565.67,719.13) node  [align=left] {1};
    \draw (437.67,718.13) node  [align=left] {1..*};
    \draw (743.07,716.13) node  [align=left] {is};
    \draw (664.27,689.73) node  [align=left] {*};
    \draw (694.07,702.93) node  [align=left] {1};
    \draw (681.55,656.93) node  [align=left] {consumes};
    \draw (297.47,873.13) node  [align=left] {mechanically linked};
    \draw (861.95,726) node  [align=left] {\ \ Software\\ \ \ component};
    \draw (628.95,727) node  [align=left] {Service};
    \draw (629.95,818) node  [align=left] { Power\\source};
    \draw (364.75,930.67) node  [align=left] {Material Entity};
    \draw (860.35,362.8) node  [align=left] {Memory};
    \draw (862.35,574.8) node  [align=left] {Processor};
    \draw (364.95,818) node  [align=left] {Device};
    \draw (365.75,727.2) node  [align=left] {Resource};
    \draw (861.75,456) node  [align=left] {Data};
    \draw (628.35,574.8) node  [align=left] {Command};
    \draw (301.35,574.6) node  [align=left] {Actuator};
    \draw (648.13,283) node  [align=left] { \ \ Environment \\\ \ \ properties};
    
    \end{tikzpicture}
    \caption{\label{object} \textbf{Structural model of a connected object} -- devices, mechanically attached to the everyday objects, carry resources that, by allowing interaction with the physical world (through sensors and actuators) and the digital world (through services and software components), transcend the primary usage of these objects.}
    \end{figure*}
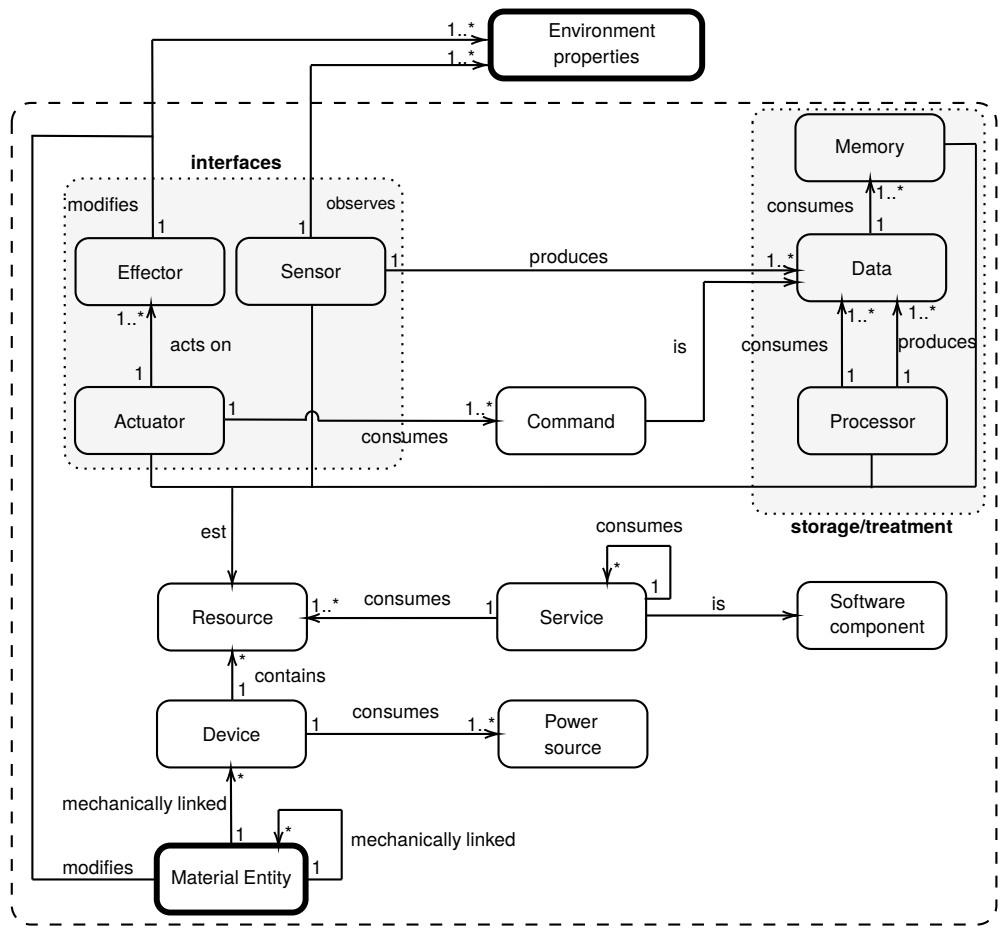

%% file: systems.tex
 
\tikzset{
pattern size/.store in=\mcSize, 
pattern size = 1pt,
pattern thickness/.store in=\mcThickness, 
pattern thickness = 0.2pt,
pattern radius/.store in=\mcRadius, 
pattern radius = 0.2pt}
\makeatletter
\pgfutil@ifundefined{pgf@pattern@name@_dv6aw9ddz}{
\makeatletter
\pgfdeclarepatternformonly[\mcRadius,\mcThickness,\mcSize]{_dv6aw9ddz}
{\pgfpoint{-0.2*\mcSize}{-0.2*\mcSize}}
{\pgfpoint{0.2*\mcSize}{0.2*\mcSize}}
{\pgfpoint{\mcSize}{\mcSize}}
{
\pgfsetcolor{\tikz@pattern@color}
\pgfsetlinewidth{\mcThickness}
\pgfpathcircle\pgfpointorigin{\mcRadius}
\pgfusepath{stroke}
}}
\makeatother

 
\tikzset{
pattern size/.store in=\mcSize, 
pattern size = 1pt,
pattern thickness/.store in=\mcThickness, 
pattern thickness = 0.2pt,
pattern radius/.store in=\mcRadius, 
pattern radius = 0.2pt}
\makeatletter
\pgfutil@ifundefined{pgf@pattern@name@_29ib4g77b}{
\makeatletter
\pgfdeclarepatternformonly[\mcRadius,\mcThickness,\mcSize]{_29ib4g77b}
{\pgfpoint{-0.2*\mcSize}{-0.2*\mcSize}}
{\pgfpoint{0.2*\mcSize}{0.2*\mcSize}}
{\pgfpoint{\mcSize}{\mcSize}}
{
\pgfsetcolor{\tikz@pattern@color}
\pgfsetlinewidth{\mcThickness}
\pgfpathcircle\pgfpointorigin{\mcRadius}
\pgfusepath{stroke}
}}
\makeatother

 
\tikzset{
pattern size/.store in=\mcSize, 
pattern size = 1pt,
pattern thickness/.store in=\mcThickness, 
pattern thickness = 0.2pt,
pattern radius/.store in=\mcRadius, 
pattern radius = 0.2pt}
\makeatletter
\pgfutil@ifundefined{pgf@pattern@name@_d27zaxjrn}{
\makeatletter
\pgfdeclarepatternformonly[\mcRadius,\mcThickness,\mcSize]{_d27zaxjrn}
{\pgfpoint{-0.2*\mcSize}{-0.2*\mcSize}}
{\pgfpoint{0.2*\mcSize}{0.2*\mcSize}}
{\pgfpoint{\mcSize}{\mcSize}}
{
\pgfsetcolor{\tikz@pattern@color}
\pgfsetlinewidth{\mcThickness}
\pgfpathcircle\pgfpointorigin{\mcRadius}
\pgfusepath{stroke}
}}
\makeatother

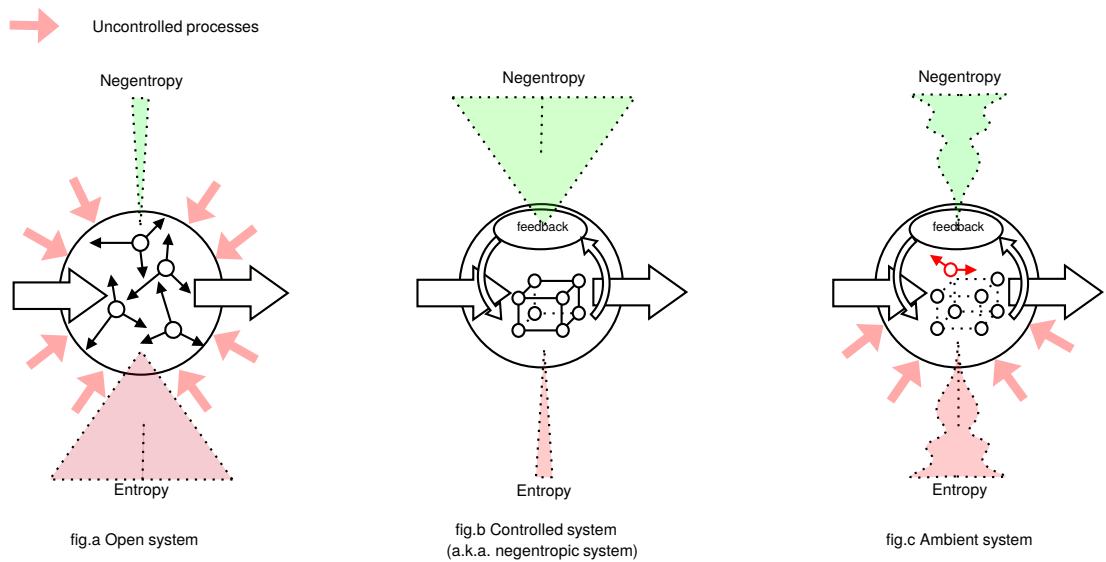
\begin{figure*}
\centering

\tikzset{every picture/.style={line width=0.75pt}} 
\begin{tikzpicture}[thick,scale=0.67, every node/.style={scale=0.67}, x=0.75pt,y=0.75pt,yscale=-1,xscale=1,font=\sffamily]

\draw  [fill={rgb, 255:red, 255; green, 255; blue, 255 }  ,fill opacity=1 ] (143.5,345) .. controls (143.5,311.03) and (171.03,283.5) .. (205,283.5) .. controls (238.97,283.5) and (266.5,311.03) .. (266.5,345) .. controls (266.5,378.97) and (238.97,406.5) .. (205,406.5) .. controls (171.03,406.5) and (143.5,378.97) .. (143.5,345) -- cycle ;
\draw  [fill={rgb, 255:red, 255; green, 255; blue, 255 }  ,fill opacity=1 ] (109,337) -- (151,337) -- (151,327) -- (179,347) -- (151,367) -- (151,357) -- (109,357) -- cycle ;
\draw  [fill={rgb, 255:red, 255; green, 255; blue, 255 }  ,fill opacity=1 ] (444.7,339.5) .. controls (444.7,305.53) and (472.23,278) .. (506.2,278) .. controls (540.17,278) and (567.7,305.53) .. (567.7,339.5) .. controls (567.7,373.47) and (540.17,401) .. (506.2,401) .. controls (472.23,401) and (444.7,373.47) .. (444.7,339.5) -- cycle ;
\draw  [fill={rgb, 255:red, 255; green, 255; blue, 255 }  ,fill opacity=1 ] (413,335) -- (455,335) -- (455,325) -- (483,345) -- (455,365) -- (455,355) -- (413,355) -- cycle ;
\draw  [fill={rgb, 255:red, 255; green, 255; blue, 255 }  ,fill opacity=1 ] (470.5,297) .. controls (470.5,288.72) and (486.17,282) .. (505.5,282) .. controls (524.83,282) and (540.5,288.72) .. (540.5,297) .. controls (540.5,305.28) and (524.83,312) .. (505.5,312) .. controls (486.17,312) and (470.5,305.28) .. (470.5,297) -- cycle ;
\draw  [fill={rgb, 255:red, 255; green, 255; blue, 255 }  ,fill opacity=1 ] (545.4,334.28) -- (587.4,334.28) -- (587.4,324.28) -- (615.4,344.28) -- (587.4,364.28) -- (587.4,354.28) -- (545.4,354.28) -- cycle ;
\draw  [fill={rgb, 255:red, 255; green, 255; blue, 255 }  ,fill opacity=1 ] (472.41,302.33) -- (472.41,302.33) .. controls (456.72,315.34) and (453.86,339.11) .. (466.01,355.44) -- (467.96,358.06) -- (464.7,360.76) -- (476.05,364.68) -- (475.13,352.12) -- (471.87,354.82) -- (469.92,352.2) .. controls (459.43,338.12) and (461.9,317.61) .. (475.43,306.39) -- (475.43,306.39) -- cycle ;
\draw  [fill={rgb, 255:red, 255; green, 255; blue, 255 }  ,fill opacity=1 ] (547.03,364.73) -- (547.03,364.73) .. controls (561.14,349.84) and (561.15,325.72) .. (547.07,310.86) -- (545.25,308.93) -- (548.18,305.83) -- (536.38,303.28) -- (538.8,315.73) -- (541.73,312.64) -- (543.56,314.56) .. controls (555.7,327.38) and (555.69,348.19) .. (543.53,361.03) -- (543.53,361.03) -- cycle ;
\draw  [fill={rgb, 255:red, 255; green, 255; blue, 255 }  ,fill opacity=1 ] (245.38,335) -- (287.38,335) -- (287.38,325) -- (315.38,345) -- (287.38,365) -- (287.38,355) -- (245.38,355) -- cycle ;
\draw  [fill={rgb, 255:red, 255; green, 0; blue, 0 }  ,fill opacity=0.2 ][dash pattern={on 0.84pt off 2.51pt}] (514.64,483.78) -- (502.16,483.78) -- (508.4,386.78) -- cycle ;
\draw  [fill={rgb, 255:red, 255; green, 255; blue, 255 }  ,fill opacity=1 ] (757.7,339.5) .. controls (757.7,305.53) and (785.23,278) .. (819.2,278) .. controls (853.17,278) and (880.7,305.53) .. (880.7,339.5) .. controls (880.7,373.47) and (853.17,401) .. (819.2,401) .. controls (785.23,401) and (757.7,373.47) .. (757.7,339.5) -- cycle ;
\draw  [fill={rgb, 255:red, 255; green, 255; blue, 255 }  ,fill opacity=1 ] (726,335) -- (768,335) -- (768,325) -- (796,345) -- (768,365) -- (768,355) -- (726,355) -- cycle ;
\draw  [fill={rgb, 255:red, 255; green, 255; blue, 255 }  ,fill opacity=1 ] (783.5,297) .. controls (783.5,288.72) and (799.17,282) .. (818.5,282) .. controls (837.83,282) and (853.5,288.72) .. (853.5,297) .. controls (853.5,305.28) and (837.83,312) .. (818.5,312) .. controls (799.17,312) and (783.5,305.28) .. (783.5,297) -- cycle ;
\draw  [fill={rgb, 255:red, 255; green, 255; blue, 255 }  ,fill opacity=1 ] (785.41,302.33) -- (785.41,302.33) .. controls (769.72,315.34) and (766.86,339.11) .. (779.01,355.44) -- (780.96,358.06) -- (777.7,360.76) -- (789.05,364.68) -- (788.13,352.12) -- (784.87,354.82) -- (782.92,352.2) .. controls (772.43,338.12) and (774.9,317.61) .. (788.43,306.39) -- (788.43,306.39) -- cycle ;
\draw  [dash pattern={on 0.84pt off 2.51pt}]  (820.24,445.4) -- (820.24,397.28) ;

\draw  [fill={rgb, 255:red, 255; green, 255; blue, 255 }  ,fill opacity=1 ] (858.4,334.28) -- (900.4,334.28) -- (900.4,324.28) -- (928.4,344.28) -- (900.4,364.28) -- (900.4,354.28) -- (858.4,354.28) -- cycle ;
\draw  [fill={rgb, 255:red, 255; green, 255; blue, 255 }  ,fill opacity=1 ] (862.03,364.73) -- (862.03,364.73) .. controls (876.14,349.84) and (876.15,325.72) .. (862.07,310.86) -- (860.25,308.93) -- (863.18,305.83) -- (851.38,303.28) -- (853.8,315.73) -- (856.73,312.64) -- (858.56,314.56) .. controls (870.7,327.38) and (870.69,348.19) .. (858.53,361.03) -- (858.53,361.03) -- cycle ;
\draw   (198.4,307.2) .. controls (198.4,303.89) and (201.09,301.2) .. (204.4,301.2) .. controls (207.71,301.2) and (210.4,303.89) .. (210.4,307.2) .. controls (210.4,310.51) and (207.71,313.2) .. (204.4,313.2) .. controls (201.09,313.2) and (198.4,310.51) .. (198.4,307.2) -- cycle ;
\draw    (167.9,307.2) -- (198.4,307.2) ;

\draw [shift={(165.9,307.2)}, rotate = 0] [fill={rgb, 255:red, 0; green, 0; blue, 0 }  ][line width=0.75]  [draw opacity=0] (8.93,-4.29) -- (0,0) -- (8.93,4.29) -- cycle    ;
\draw    (206.66,332.21) -- (204.4,313.2) ;

\draw [shift={(206.9,334.2)}, rotate = 263.21] [fill={rgb, 255:red, 0; green, 0; blue, 0 }  ][line width=0.75]  [draw opacity=0] (8.93,-4.29) -- (0,0) -- (8.93,4.29) -- cycle    ;
\draw    (221.51,289.64) -- (208.4,303.2) ;

\draw [shift={(222.9,288.2)}, rotate = 134.03] [fill={rgb, 255:red, 0; green, 0; blue, 0 }  ][line width=0.75]  [draw opacity=0] (8.93,-4.29) -- (0,0) -- (8.93,4.29) -- cycle    ;
\draw   (182.79,362) .. controls (180.13,360.02) and (179.58,356.26) .. (181.57,353.6) .. controls (183.55,350.95) and (187.31,350.4) .. (189.96,352.38) .. controls (192.62,354.37) and (193.17,358.13) .. (191.18,360.78) .. controls (189.2,363.44) and (185.44,363.98) .. (182.79,362) -- cycle ;
\draw    (164.55,386.45) -- (182.79,362) ;

\draw [shift={(163.35,388.05)}, rotate = 306.73] [fill={rgb, 255:red, 0; green, 0; blue, 0 }  ][line width=0.75]  [draw opacity=0] (8.93,-4.29) -- (0,0) -- (8.93,4.29) -- cycle    ;
\draw    (207.78,370.34) -- (191.18,360.78) ;

\draw [shift={(209.51,371.34)}, rotate = 209.94] [fill={rgb, 255:red, 0; green, 0; blue, 0 }  ][line width=0.75]  [draw opacity=0] (8.93,-4.29) -- (0,0) -- (8.93,4.29) -- cycle    ;
\draw    (182.53,332.98) -- (185.56,351.6) ;

\draw [shift={(182.21,331)}, rotate = 80.76] [fill={rgb, 255:red, 0; green, 0; blue, 0 }  ][line width=0.75]  [draw opacity=0] (8.93,-4.29) -- (0,0) -- (8.93,4.29) -- cycle    ;

\draw   (218.96,329.97) .. controls (216.86,327.41) and (217.24,323.63) .. (219.81,321.53) .. controls (222.37,319.43) and (226.15,319.81) .. (228.25,322.37) .. controls (230.35,324.94) and (229.97,328.72) .. (227.41,330.82) .. controls (224.84,332.91) and (221.06,332.54) .. (218.96,329.97) -- cycle ;
\draw    (195.36,349.29) -- (218.96,329.97) ;

\draw [shift={(193.81,350.55)}, rotate = 320.71] [fill={rgb, 255:red, 0; green, 0; blue, 0 }  ][line width=0.75]  [draw opacity=0] (8.93,-4.29) -- (0,0) -- (8.93,4.29) -- cycle    ;
\draw    (241.2,344.1) -- (227.41,330.82) ;

\draw [shift={(242.64,345.49)}, rotate = 223.92] [fill={rgb, 255:red, 0; green, 0; blue, 0 }  ][line width=0.75]  [draw opacity=0] (8.93,-4.29) -- (0,0) -- (8.93,4.29) -- cycle    ;
\draw    (225.73,301.74) -- (224.17,320.54) ;

\draw [shift={(225.89,299.75)}, rotate = 94.74] [fill={rgb, 255:red, 0; green, 0; blue, 0 }  ][line width=0.75]  [draw opacity=0] (8.93,-4.29) -- (0,0) -- (8.93,4.29) -- cycle    ;

\draw   (227.4,367.05) .. controls (230.57,366.08) and (233.92,367.86) .. (234.89,371.03) .. controls (235.87,374.19) and (234.09,377.55) .. (230.92,378.52) .. controls (227.75,379.49) and (224.39,377.71) .. (223.42,374.55) .. controls (222.45,371.38) and (224.23,368.02) .. (227.4,367.05) -- cycle ;
\draw    (218.45,337.89) -- (227.4,367.05) ;

\draw [shift={(217.86,335.98)}, rotate = 72.94] [fill={rgb, 255:red, 0; green, 0; blue, 0 }  ][line width=0.75]  [draw opacity=0] (8.93,-4.29) -- (0,0) -- (8.93,4.29) -- cycle    ;
\draw    (205.91,382.29) -- (223.42,374.55) ;

\draw [shift={(204.08,383.1)}, rotate = 336.15] [fill={rgb, 255:red, 0; green, 0; blue, 0 }  ][line width=0.75]  [draw opacity=0] (8.93,-4.29) -- (0,0) -- (8.93,4.29) -- cycle    ;
\draw    (250.97,383.99) -- (234.16,375.44) ;

\draw [shift={(252.75,384.9)}, rotate = 206.97] [fill={rgb, 255:red, 0; green, 0; blue, 0 }  ][line width=0.75]  [draw opacity=0] (8.93,-4.29) -- (0,0) -- (8.93,4.29) -- cycle    ;

\draw   (489.04,348.83) -- (502.01,335.86) -- (534.46,335.86) -- (534.46,360.17) -- (521.49,373.14) -- (489.04,373.14) -- cycle ; \draw   (534.46,335.86) -- (521.49,348.83) -- (489.04,348.83) ; \draw   (521.49,348.83) -- (521.49,373.14) ;
\draw  [fill={rgb, 255:red, 255; green, 255; blue, 255 }  ,fill opacity=1 ] (484.5,348.83) .. controls (484.5,346.14) and (486.53,343.97) .. (489.04,343.97) .. controls (491.55,343.97) and (493.58,346.14) .. (493.58,348.83) .. controls (493.58,351.51) and (491.55,353.69) .. (489.04,353.69) .. controls (486.53,353.69) and (484.5,351.51) .. (484.5,348.83) -- cycle ;
\draw  [fill={rgb, 255:red, 255; green, 255; blue, 255 }  ,fill opacity=1 ] (517.81,348.83) .. controls (517.81,346.14) and (519.84,343.97) .. (522.35,343.97) .. controls (524.86,343.97) and (526.89,346.14) .. (526.89,348.83) .. controls (526.89,351.51) and (524.86,353.69) .. (522.35,353.69) .. controls (519.84,353.69) and (517.81,351.51) .. (517.81,348.83) -- cycle ;
\draw  [fill={rgb, 255:red, 255; green, 255; blue, 255 }  ,fill opacity=1 ] (529.92,335.86) .. controls (529.92,333.18) and (531.95,331) .. (534.46,331) .. controls (536.97,331) and (539,333.18) .. (539,335.86) .. controls (539,338.55) and (536.97,340.72) .. (534.46,340.72) .. controls (531.95,340.72) and (529.92,338.55) .. (529.92,335.86) -- cycle ;
\draw  [fill={rgb, 255:red, 255; green, 255; blue, 255 }  ,fill opacity=1 ] (529.92,360.17) .. controls (529.92,357.49) and (531.95,355.31) .. (534.46,355.31) .. controls (536.97,355.31) and (539,357.49) .. (539,360.17) .. controls (539,362.86) and (536.97,365.03) .. (534.46,365.03) .. controls (531.95,365.03) and (529.92,362.86) .. (529.92,360.17) -- cycle ;
\draw  [fill={rgb, 255:red, 255; green, 255; blue, 255 }  ,fill opacity=1 ] (517.81,373.14) .. controls (517.81,370.45) and (519.84,368.28) .. (522.35,368.28) .. controls (524.86,368.28) and (526.89,370.45) .. (526.89,373.14) .. controls (526.89,375.82) and (524.86,378) .. (522.35,378) .. controls (519.84,378) and (517.81,375.82) .. (517.81,373.14) -- cycle ;
\draw  [dash pattern={on 0.84pt off 2.51pt}]  (501.28,360.4) -- (489.04,373.14) ;

\draw  [dash pattern={on 0.84pt off 2.51pt}]  (501.28,360.4) -- (534.46,360.17) ;

\draw  [dash pattern={on 0.84pt off 2.51pt}]  (501.28,360.4) -- (501.15,335.86) ;

\draw  [fill={rgb, 255:red, 255; green, 255; blue, 255 }  ,fill opacity=1 ] (496.61,335.86) .. controls (496.61,333.18) and (498.64,331) .. (501.15,331) .. controls (503.66,331) and (505.69,333.18) .. (505.69,335.86) .. controls (505.69,338.55) and (503.66,340.72) .. (501.15,340.72) .. controls (498.64,340.72) and (496.61,338.55) .. (496.61,335.86) -- cycle ;
\draw  [fill={rgb, 255:red, 255; green, 255; blue, 255 }  ,fill opacity=1 ] (496.74,360.4) .. controls (496.74,357.71) and (498.77,355.54) .. (501.28,355.54) .. controls (503.79,355.54) and (505.82,357.71) .. (505.82,360.4) .. controls (505.82,363.08) and (503.79,365.26) .. (501.28,365.26) .. controls (498.77,365.26) and (496.74,363.08) .. (496.74,360.4) -- cycle ;
\draw  [fill={rgb, 255:red, 255; green, 255; blue, 255 }  ,fill opacity=1 ] (484.5,373.14) .. controls (484.5,370.45) and (486.53,368.28) .. (489.04,368.28) .. controls (491.55,368.28) and (493.58,370.45) .. (493.58,373.14) .. controls (493.58,375.82) and (491.55,378) .. (489.04,378) .. controls (486.53,378) and (484.5,375.82) .. (484.5,373.14) -- cycle ;

\draw  [dash pattern={on 0.84pt off 2.51pt}] (804.24,347.43) -- (817.21,334.46) -- (849.66,334.46) -- (849.66,358.77) -- (836.69,371.74) -- (804.24,371.74) -- cycle ; \draw  [dash pattern={on 0.84pt off 2.51pt}] (849.66,334.46) -- (836.69,347.43) -- (804.24,347.43) ; \draw  [dash pattern={on 0.84pt off 2.51pt}] (836.69,347.43) -- (836.69,371.74) ;
\draw  [fill={rgb, 255:red, 255; green, 255; blue, 255 }  ,fill opacity=1 ] (799.7,347.43) .. controls (799.7,344.74) and (801.73,342.57) .. (804.24,342.57) .. controls (806.75,342.57) and (808.78,344.74) .. (808.78,347.43) .. controls (808.78,350.11) and (806.75,352.29) .. (804.24,352.29) .. controls (801.73,352.29) and (799.7,350.11) .. (799.7,347.43) -- cycle ;
\draw  [fill={rgb, 255:red, 255; green, 255; blue, 255 }  ,fill opacity=1 ] (833.01,347.43) .. controls (833.01,344.74) and (835.04,342.57) .. (837.55,342.57) .. controls (840.06,342.57) and (842.09,344.74) .. (842.09,347.43) .. controls (842.09,350.11) and (840.06,352.29) .. (837.55,352.29) .. controls (835.04,352.29) and (833.01,350.11) .. (833.01,347.43) -- cycle ;
\draw  [fill={rgb, 255:red, 255; green, 255; blue, 255 }  ,fill opacity=1 ] (845.12,334.46) .. controls (845.12,331.78) and (847.15,329.6) .. (849.66,329.6) .. controls (852.17,329.6) and (854.2,331.78) .. (854.2,334.46) .. controls (854.2,337.15) and (852.17,339.32) .. (849.66,339.32) .. controls (847.15,339.32) and (845.12,337.15) .. (845.12,334.46) -- cycle ;
\draw  [fill={rgb, 255:red, 255; green, 255; blue, 255 }  ,fill opacity=1 ] (833.01,371.74) .. controls (833.01,369.05) and (835.04,366.88) .. (837.55,366.88) .. controls (840.06,366.88) and (842.09,369.05) .. (842.09,371.74) .. controls (842.09,374.42) and (840.06,376.6) .. (837.55,376.6) .. controls (835.04,376.6) and (833.01,374.42) .. (833.01,371.74) -- cycle ;
\draw  [dash pattern={on 0.84pt off 2.51pt}]  (816.48,359) -- (804.24,371.74) ;

\draw  [dash pattern={on 0.84pt off 2.51pt}]  (816.48,359) -- (849.66,358.77) ;

\draw  [dash pattern={on 0.84pt off 2.51pt}]  (817.48,359) -- (817.21,334.46) ;

\draw  [color={rgb, 255:red, 255; green, 0; blue, 0 }  ,draw opacity=1 ][fill={rgb, 255:red, 255; green, 255; blue, 255 }  ,fill opacity=1 ] (809.81,327.46) .. controls (809.81,324.78) and (811.84,322.6) .. (814.35,322.6) .. controls (816.86,322.6) and (818.89,324.78) .. (818.89,327.46) .. controls (818.89,330.15) and (816.86,332.32) .. (814.35,332.32) .. controls (811.84,332.32) and (809.81,330.15) .. (809.81,327.46) -- cycle ;
\draw  [fill={rgb, 255:red, 255; green, 255; blue, 255 }  ,fill opacity=1 ] (812.94,359) .. controls (812.94,356.31) and (814.97,354.14) .. (817.48,354.14) .. controls (819.99,354.14) and (822.02,356.31) .. (822.02,359) .. controls (822.02,361.68) and (819.99,363.86) .. (817.48,363.86) .. controls (814.97,363.86) and (812.94,361.68) .. (812.94,359) -- cycle ;
\draw  [fill={rgb, 255:red, 255; green, 255; blue, 255 }  ,fill opacity=1 ] (799.7,371.74) .. controls (799.7,369.05) and (801.73,366.88) .. (804.24,366.88) .. controls (806.75,366.88) and (808.78,369.05) .. (808.78,371.74) .. controls (808.78,374.42) and (806.75,376.6) .. (804.24,376.6) .. controls (801.73,376.6) and (799.7,374.42) .. (799.7,371.74) -- cycle ;
\draw [color={rgb, 255:red, 255; green, 0; blue, 0 }  ,draw opacity=1 ]   (800.35,317.33) -- (810.81,324.46) ;

\draw [shift={(798.7,316.2)}, rotate = 34.3] [fill={rgb, 255:red, 255; green, 0; blue, 0 }  ,fill opacity=1 ][line width=0.75]  [draw opacity=0] (8.93,-4.29) -- (0,0) -- (8.93,4.29) -- cycle    ;
\draw [color={rgb, 255:red, 255; green, 0; blue, 0 }  ,draw opacity=1 ]   (832.3,327.58) -- (818.89,327.46) ;

\draw [shift={(834.3,327.6)}, rotate = 180.51] [fill={rgb, 255:red, 255; green, 0; blue, 0 }  ,fill opacity=1 ][line width=0.75]  [draw opacity=0] (8.93,-4.29) -- (0,0) -- (8.93,4.29) -- cycle    ;
\draw  [fill={rgb, 255:red, 255; green, 255; blue, 255 }  ,fill opacity=1 ] (845.12,358.77) .. controls (845.12,356.09) and (847.15,353.91) .. (849.66,353.91) .. controls (852.17,353.91) and (854.2,356.09) .. (854.2,358.77) .. controls (854.2,361.46) and (852.17,363.63) .. (849.66,363.63) .. controls (847.15,363.63) and (845.12,361.46) .. (845.12,358.77) -- cycle ;
\draw  [fill={rgb, 255:red, 36; green, 255; blue, 0 }  ,fill opacity=0.2 ][dash pattern={on 0.84pt off 2.51pt}] (439.25,197.5) -- (576.75,197.5) -- (508,294.5) -- cycle ;
\draw [fill={rgb, 255:red, 36; green, 255; blue, 0 }  ,fill opacity=0.2 ] [dash pattern={on 0.84pt off 2.51pt}]  (507,197.5) -- (506,240.63) ;

\draw  [fill={rgb, 255:red, 0; green, 255; blue, 8 }  ,fill opacity=0.2 ][dash pattern={on 0.84pt off 2.51pt}] (821.47,195.85) .. controls (829.53,196.15) and (856.93,193.74) .. (856.49,195.93) .. controls (856.05,198.12) and (830.6,203.41) .. (841.56,213.83) .. controls (852.52,224.25) and (820.1,223.95) .. (831.13,234.98) .. controls (842.17,246.01) and (832.12,256.28) .. (825.88,267.15) .. controls (819.64,278.03) and (820.25,296.91) .. (819.87,298.5) .. controls (819.49,300.09) and (819.94,278.03) .. (814.54,266.55) .. controls (809.14,255.07) and (798.48,246.01) .. (809.14,234.83) .. controls (819.79,223.65) and (787.45,225.16) .. (798.71,214.43) .. controls (809.97,203.71) and (783.83,196.91) .. (783.24,195.93) .. controls (782.65,194.95) and (813.4,195.55) .. (821.47,195.85) -- cycle ;
\draw  [fill={rgb, 255:red, 255; green, 0; blue, 5 }  ,fill opacity=0.2 ][dash pattern={on 0.84pt off 2.51pt}] (818.27,483.15) .. controls (810.2,482.85) and (782.81,485.26) .. (783.24,483.07) .. controls (783.68,480.88) and (809.14,475.59) .. (798.18,465.17) .. controls (787.22,454.75) and (819.64,455.05) .. (808.6,444.02) .. controls (797.57,432.99) and (807.61,422.72) .. (813.86,411.85) .. controls (820.1,400.97) and (819.49,382.09) .. (819.87,380.5) .. controls (820.25,378.91) and (819.79,400.97) .. (825.19,412.45) .. controls (830.6,423.93) and (841.25,432.99) .. (830.6,444.17) .. controls (819.94,455.35) and (852.29,453.84) .. (841.02,464.57) .. controls (829.76,475.29) and (855.9,482.09) .. (856.49,483.07) .. controls (857.08,484.05) and (826.34,483.45) .. (818.27,483.15) -- cycle ;
\draw  [fill={rgb, 255:red, 0; green, 255; blue, 10 }  ,fill opacity=0.2 ][dash pattern={on 0.84pt off 2.51pt}] (198.16,198.2) -- (210.64,198.2) -- (204.4,295.2) -- cycle ;
\draw  [fill={rgb, 255:red, 208; green, 2; blue, 27 }  ,fill opacity=0.2 ][dash pattern={on 0.84pt off 2.51pt}] (273.75,485.5) -- (136.25,485.5) -- (205,388.5) -- cycle ;
\draw [fill={rgb, 255:red, 208; green, 2; blue, 27 }  ,fill opacity=0.2 ] [dash pattern={on 0.84pt off 2.51pt}]  (206,485.5) -- (207,442.38) ;

\draw  [draw opacity=0][fill={rgb, 255:red, 255; green, 0; blue, 0 }  ,fill opacity=0.34 ] (290.28,398.14) -- (271.59,387.52) -- (266.59,396.33) -- (259.34,376.69) -- (279.92,372.85) -- (274.92,381.65) -- (293.62,392.27) -- cycle ;
\draw  [draw opacity=0][fill={rgb, 255:red, 255; green, 0; blue, 0 }  ,fill opacity=0.34 ] (252.92,435.1) -- (240.3,417.7) -- (232.1,423.64) -- (233.64,402.77) -- (253.96,407.79) -- (245.76,413.74) -- (258.38,431.14) -- cycle ;
\draw  [draw opacity=0][fill={rgb, 255:red, 255; green, 0; blue, 0 }  ,fill opacity=0.34 ] (118.14,397.48) -- (135.1,384.27) -- (128.88,376.28) -- (149.8,377.1) -- (145.47,397.58) -- (139.25,389.59) -- (122.29,402.8) -- cycle ;
\draw  [draw opacity=0][fill={rgb, 255:red, 255; green, 0; blue, 0 }  ,fill opacity=0.34 ] (152.22,432.46) -- (164.43,414.76) -- (156.1,409.01) -- (176.29,403.51) -- (178.32,424.35) -- (169.98,418.6) -- (157.77,436.29) -- cycle ;

\draw  [draw opacity=0][fill={rgb, 255:red, 255; green, 0; blue, 0 }  ,fill opacity=0.34 ] (119.11,295.31) -- (137.81,305.92) -- (142.81,297.12) -- (150.05,316.76) -- (129.48,320.6) -- (134.47,311.79) -- (115.78,301.18) -- cycle ;
\draw  [draw opacity=0][fill={rgb, 255:red, 255; green, 0; blue, 0 }  ,fill opacity=0.34 ] (156.65,257.14) -- (165.98,276.51) -- (175.1,272.11) -- (169.89,292.39) -- (150.78,283.84) -- (159.9,279.44) -- (150.57,260.08) -- cycle ;
\draw  [draw opacity=0][fill={rgb, 255:red, 255; green, 0; blue, 0 }  ,fill opacity=0.34 ] (292.62,298.64) -- (275.65,311.85) -- (281.88,319.84) -- (260.96,319.02) -- (265.28,298.54) -- (271.51,306.53) -- (288.47,293.31) -- cycle ;
\draw  [draw opacity=0][fill={rgb, 255:red, 255; green, 0; blue, 0 }  ,fill opacity=0.34 ] (265.18,263.99) -- (252.97,281.68) -- (261.3,287.43) -- (241.1,292.93) -- (239.08,272.1) -- (247.41,277.85) -- (259.62,260.15) -- cycle ;

\draw  [draw opacity=0][fill={rgb, 255:red, 255; green, 0; blue, 0 }  ,fill opacity=0.34 ] (904.28,390.14) -- (885.59,379.52) -- (880.59,388.33) -- (873.34,368.69) -- (893.92,364.85) -- (888.92,373.65) -- (907.62,384.27) -- cycle ;
\draw  [draw opacity=0][fill={rgb, 255:red, 255; green, 0; blue, 0 }  ,fill opacity=0.34 ] (866.92,427.1) -- (854.3,409.7) -- (846.1,415.64) -- (847.64,394.77) -- (867.96,399.79) -- (859.76,405.74) -- (872.38,423.14) -- cycle ;
\draw  [draw opacity=0][fill={rgb, 255:red, 255; green, 0; blue, 0 }  ,fill opacity=0.34 ] (732.14,389.48) -- (749.1,376.27) -- (742.88,368.28) -- (763.8,369.1) -- (759.47,389.58) -- (753.25,381.59) -- (736.29,394.8) -- cycle ;
\draw  [draw opacity=0][fill={rgb, 255:red, 255; green, 0; blue, 0 }  ,fill opacity=0.34 ] (766.22,424.46) -- (778.43,406.76) -- (770.1,401.01) -- (790.29,395.51) -- (792.32,416.35) -- (783.98,410.6) -- (771.77,428.29) -- cycle ;

\draw  [draw opacity=0][fill={rgb, 255:red, 255; green, 0; blue, 0 }  ,fill opacity=0.34 ] (106.08,139.73) -- (127.57,140.18) -- (127.78,130.06) -- (143.5,143.9) -- (127.21,157.06) -- (127.43,146.93) -- (105.93,146.48) -- cycle ;
\draw (231,145) node  [align=left] {Uncontrolled processes};

\draw (200,532) node  [align=left] {fig.a Open system};
\draw (507,532) node  [align=left] {\ \ fig.b Controlled system \\ (a.k.a. negentropic system)};
\draw (507.41,294.78) node  [align=left] {{\footnotesize feedback}};
\draw (205,494) node  [align=left] {Entropy};
\draw (820.41,294.78) node  [align=left] {{\footnotesize feedback}};
\draw (821,532) node  [align=left] {fig.c Ambient system};
\draw (205,186) node  [align=left] {Negentropy};
\draw (508,495) node  [align=left] {Entropy};
\draw (508,184) node  [align=left] {Negentropy};
\draw (821,494) node  [align=left] {Entropy};
\draw (821,183) node  [align=left] {Negentropy};

\end{tikzpicture}

\caption{\label{systems}Open, controlled and ambient systems, entropy and neguentropy. }
\end{figure*}

%% file: contextual.tex
\begin{figure*}
    \centering
    
    \tikzset{every picture/.style={line width=0.75pt}} 
    \begin{tikzpicture}[thick,scale=0.67, every node/.style={scale=0.67}, x=0.75pt,y=0.75pt,yscale=-1,xscale=1,font=\sffamily]
    
        \draw  [color={rgb, 255:red, 189; green, 16; blue, 224 }  ,draw opacity=1 ][fill={rgb, 255:red, 189; green, 16; blue, 224 }  ,fill opacity=1 ][line width=1.5]  (339.77,159.11) .. controls (339.77,156.05) and (342.3,153.56) .. (345.42,153.56) .. controls (348.54,153.56) and (351.06,156.05) .. (351.06,159.11) .. controls (351.06,162.17) and (348.54,164.65) .. (345.42,164.65) .. controls (342.3,164.65) and (339.77,162.17) .. (339.77,159.11) -- cycle ;
        \draw  [color={rgb, 255:red, 189; green, 16; blue, 224 }  ,draw opacity=1 ][dash pattern={on 1.69pt off 2.76pt}][line width=1.5]  (333.39,159.11) .. controls (333.39,152.58) and (338.77,147.29) .. (345.42,147.29) .. controls (352.06,147.29) and (357.45,152.58) .. (357.45,159.11) .. controls (357.45,165.64) and (352.06,170.93) .. (345.42,170.93) .. controls (338.77,170.93) and (333.39,165.64) .. (333.39,159.11) -- cycle ;
        \draw  [color={rgb, 255:red, 189; green, 16; blue, 224 }  ,draw opacity=1 ][dash pattern={on 1.69pt off 2.76pt}][line width=1.5]  (322.93,159.11) .. controls (322.93,146.91) and (333,137.02) .. (345.42,137.02) .. controls (357.84,137.02) and (367.9,146.91) .. (367.9,159.11) .. controls (367.9,171.31) and (357.84,181.2) .. (345.42,181.2) .. controls (333,181.2) and (322.93,171.31) .. (322.93,159.11) -- cycle ;
        \draw  [color={rgb, 255:red, 189; green, 16; blue, 224 }  ,draw opacity=1 ][dash pattern={on 1.69pt off 2.76pt}][line width=1.5]  (309.37,159.11) .. controls (309.37,139.55) and (325.51,123.7) .. (345.42,123.7) .. controls (365.33,123.7) and (381.47,139.55) .. (381.47,159.11) .. controls (381.47,178.67) and (365.33,194.52) .. (345.42,194.52) .. controls (325.51,194.52) and (309.37,178.67) .. (309.37,159.11) -- cycle ;
        \draw  [color={rgb, 255:red, 189; green, 16; blue, 224 }  ,draw opacity=1 ][dash pattern={on 1.69pt off 2.76pt}][line width=1.5]  (284.41,159.11) .. controls (284.41,126.01) and (311.73,99.18) .. (345.42,99.18) .. controls (379.11,99.18) and (406.42,126.01) .. (406.42,159.11) .. controls (406.42,192.21) and (379.11,219.04) .. (345.42,219.04) .. controls (311.73,219.04) and (284.41,192.21) .. (284.41,159.11) -- cycle ;
        \draw  [color={rgb, 255:red, 189; green, 16; blue, 224 }  ,draw opacity=1 ][dash pattern={on 1.69pt off 2.76pt}][line width=1.5]  (253.52,159.11) .. controls (253.52,109.25) and (294.66,68.83) .. (345.42,68.83) .. controls (396.17,68.83) and (437.32,109.25) .. (437.32,159.11) .. controls (437.32,208.97) and (396.17,249.39) .. (345.42,249.39) .. controls (294.66,249.39) and (253.52,208.97) .. (253.52,159.11) -- cycle ;
        
        \draw  [fill={rgb, 255:red, 155; green, 155; blue, 155 }  ,fill opacity=0.1 ][line width=1.5]  (405.81,165.33) .. controls (405.81,129.07) and (435.74,99.67) .. (472.66,99.67) .. controls (509.57,99.67) and (539.5,129.07) .. (539.5,165.33) .. controls (539.5,201.6) and (509.57,231) .. (472.66,231) .. controls (435.74,231) and (405.81,201.6) .. (405.81,165.33) -- cycle ;
        \draw [line width=2.25]  [dash pattern={on 6.75pt off 4.5pt}]  (461.21,154.32) .. controls (413.73,71.63) and (403.59,290.01) .. (346.29,166.54) ;
        \draw [shift={(345.42,164.65)}, rotate = 425.52] [color={rgb, 255:red, 0; green, 0; blue, 0 }  ][line width=2.25]    (17.49,-5.26) .. controls (11.12,-2.23) and (5.29,-0.48) .. (0,0) .. controls (5.29,0.48) and (11.12,2.23) .. (17.49,5.26)   ;
        
        \draw  [fill={rgb, 255:red, 155; green, 155; blue, 155 }  ,fill opacity=0.1 ][line width=1.5]  (68.84,166) .. controls (68.84,129.45) and (98.63,99.81) .. (135.39,99.81) .. controls (172.14,99.81) and (201.94,129.45) .. (201.94,166) .. controls (201.94,202.55) and (172.14,232.19) .. (135.39,232.19) .. controls (98.63,232.19) and (68.84,202.55) .. (68.84,166) -- cycle ;
        \draw  [color={rgb, 255:red, 189; green, 16; blue, 224 }  ,draw opacity=1 ][fill={rgb, 255:red, 189; green, 16; blue, 224 }  ,fill opacity=1 ] (107.38,166) .. controls (107.38,162.91) and (109.9,160.41) .. (113,160.41) .. controls (116.1,160.41) and (118.62,162.91) .. (118.62,166) .. controls (118.62,169.09) and (116.1,171.59) .. (113,171.59) .. controls (109.9,171.59) and (107.38,169.09) .. (107.38,166) -- cycle ;
        \draw  [color={rgb, 255:red, 189; green, 16; blue, 224 }  ,draw opacity=1 ][dash pattern={on 1.69pt off 2.76pt}][line width=1.5]  (101.02,166) .. controls (101.02,159.42) and (106.38,154.09) .. (113,154.09) .. controls (119.62,154.09) and (124.98,159.42) .. (124.98,166) .. controls (124.98,172.58) and (119.62,177.91) .. (113,177.91) .. controls (106.38,177.91) and (101.02,172.58) .. (101.02,166) -- cycle ;
        \draw  [color={rgb, 255:red, 189; green, 16; blue, 224 }  ,draw opacity=1 ][dash pattern={on 1.69pt off 2.76pt}][line width=1.5]  (90.61,166) .. controls (90.61,153.7) and (100.64,143.73) .. (113,143.73) .. controls (125.36,143.73) and (135.39,153.7) .. (135.39,166) .. controls (135.39,178.3) and (125.36,188.27) .. (113,188.27) .. controls (100.64,188.27) and (90.61,178.3) .. (90.61,166) -- cycle ;
        \draw  [color={rgb, 255:red, 189; green, 16; blue, 224 }  ,draw opacity=1 ][dash pattern={on 1.69pt off 2.76pt}][line width=1.5]  (77.11,166) .. controls (77.11,146.29) and (93.18,130.3) .. (113,130.3) .. controls (132.82,130.3) and (148.89,146.29) .. (148.89,166) .. controls (148.89,185.71) and (132.82,201.7) .. (113,201.7) .. controls (93.18,201.7) and (77.11,185.71) .. (77.11,166) -- cycle ;
        \draw  [color={rgb, 255:red, 189; green, 16; blue, 224 }  ,draw opacity=1 ][dash pattern={on 1.69pt off 2.76pt}][line width=1.5]  (52.26,166) .. controls (52.26,132.64) and (79.46,105.59) .. (113,105.59) .. controls (146.54,105.59) and (173.74,132.64) .. (173.74,166) .. controls (173.74,199.36) and (146.54,226.41) .. (113,226.41) .. controls (79.46,226.41) and (52.26,199.36) .. (52.26,166) -- cycle ;
        \draw  [color={rgb, 255:red, 189; green, 16; blue, 224 }  ,draw opacity=1 ][dash pattern={on 1.69pt off 2.76pt}][line width=1.5]  (21.5,166) .. controls (21.5,115.74) and (62.47,75) .. (113,75) .. controls (163.53,75) and (204.5,115.74) .. (204.5,166) .. controls (204.5,216.26) and (163.53,257) .. (113,257) .. controls (62.47,257) and (21.5,216.26) .. (21.5,166) -- cycle ;
        \draw  [fill={rgb, 255:red, 155; green, 155; blue, 155 }  ,fill opacity=0.21 ] (230,39.5) .. controls (230,30.94) and (236.94,24) .. (245.5,24) .. controls (254.06,24) and (261,30.94) .. (261,39.5) .. controls (261,48.06) and (254.06,55) .. (245.5,55) .. controls (236.94,55) and (230,48.06) .. (230,39.5) -- cycle ;
        \draw  [color={rgb, 255:red, 189; green, 16; blue, 224 }  ,draw opacity=1 ][fill={rgb, 255:red, 189; green, 16; blue, 224 }  ,fill opacity=1 ][line width=1.5]  (24.58,40.43) .. controls (24.58,38.97) and (25.8,37.78) .. (27.31,37.78) .. controls (28.81,37.78) and (30.03,38.97) .. (30.03,40.43) .. controls (30.03,41.88) and (28.81,43.07) .. (27.31,43.07) .. controls (25.8,43.07) and (24.58,41.88) .. (24.58,40.43) -- cycle ;
        \draw  [color={rgb, 255:red, 189; green, 16; blue, 224 }  ,draw opacity=1 ][dash pattern={on 1.69pt off 2.76pt}][line width=1.5]  (19.06,40.43) .. controls (19.06,36.01) and (22.75,32.43) .. (27.31,32.43) .. controls (31.86,32.43) and (35.55,36.01) .. (35.55,40.43) .. controls (35.55,44.84) and (31.86,48.42) .. (27.31,48.42) .. controls (22.75,48.42) and (19.06,44.84) .. (19.06,40.43) -- cycle ;
        \draw  [color={rgb, 255:red, 189; green, 16; blue, 224 }  ,draw opacity=1 ][dash pattern={on 1.69pt off 2.76pt}][line width=1.5]  (11.24,40.43) .. controls (11.24,31.82) and (18.43,24.85) .. (27.31,24.85) .. controls (36.18,24.85) and (43.38,31.82) .. (43.38,40.43) .. controls (43.38,49.03) and (36.18,56) .. (27.31,56) .. controls (18.43,56) and (11.24,49.03) .. (11.24,40.43) -- cycle ;
        
        \draw  [color={rgb, 255:red, 208; green, 2; blue, 27 }  ,draw opacity=1 ][fill={rgb, 255:red, 208; green, 2; blue, 27 }  ,fill opacity=1 ][line width=1.5]  (95.14,39.87) .. controls (95.14,38.41) and (96.36,37.23) .. (97.87,37.23) .. controls (99.37,37.23) and (100.59,38.41) .. (100.59,39.87) .. controls (100.59,41.33) and (99.37,42.51) .. (97.87,42.51) .. controls (96.36,42.51) and (95.14,41.33) .. (95.14,39.87) -- cycle ;
        \draw  [color={rgb, 255:red, 208; green, 2; blue, 27 }  ,draw opacity=1 ][dash pattern={on 1.69pt off 2.76pt}][line width=1.5]  (89.62,39.87) .. controls (89.62,35.45) and (93.31,31.88) .. (97.87,31.88) .. controls (102.42,31.88) and (106.11,35.45) .. (106.11,39.87) .. controls (106.11,44.28) and (102.42,47.86) .. (97.87,47.86) .. controls (93.31,47.86) and (89.62,44.28) .. (89.62,39.87) -- cycle ;
        \draw  [color={rgb, 255:red, 208; green, 2; blue, 27 }  ,draw opacity=1 ][dash pattern={on 1.69pt off 2.76pt}][line width=1.5]  (81.8,39.87) .. controls (81.8,31.27) and (88.99,24.29) .. (97.87,24.29) .. controls (106.74,24.29) and (113.93,31.27) .. (113.93,39.87) .. controls (113.93,48.47) and (106.74,55.44) .. (97.87,55.44) .. controls (88.99,55.44) and (81.8,48.47) .. (81.8,39.87) -- cycle ;
        \draw  [color={rgb, 255:red, 74; green, 144; blue, 226 }  ,draw opacity=1 ][fill={rgb, 255:red, 74; green, 144; blue, 226 }  ,fill opacity=1 ][line width=1.5]  (60.15,39.87) .. controls (60.15,38.41) and (61.37,37.23) .. (62.87,37.23) .. controls (64.38,37.23) and (65.6,38.41) .. (65.6,39.87) .. controls (65.6,41.33) and (64.38,42.51) .. (62.87,42.51) .. controls (61.37,42.51) and (60.15,41.33) .. (60.15,39.87) -- cycle ;
        \draw  [color={rgb, 255:red, 74; green, 144; blue, 226 }  ,draw opacity=1 ][dash pattern={on 1.69pt off 2.76pt}][line width=1.5]  (54.63,39.87) .. controls (54.63,35.45) and (58.32,31.88) .. (62.87,31.88) .. controls (67.43,31.88) and (71.12,35.45) .. (71.12,39.87) .. controls (71.12,44.28) and (67.43,47.86) .. (62.87,47.86) .. controls (58.32,47.86) and (54.63,44.28) .. (54.63,39.87) -- cycle ;
        \draw  [color={rgb, 255:red, 74; green, 144; blue, 226 }  ,draw opacity=1 ][dash pattern={on 1.69pt off 2.76pt}][line width=1.5]  (46.8,39.87) .. controls (46.8,31.27) and (54,24.29) .. (62.87,24.29) .. controls (71.75,24.29) and (78.94,31.27) .. (78.94,39.87) .. controls (78.94,48.47) and (71.75,55.44) .. (62.87,55.44) .. controls (54,55.44) and (46.8,48.47) .. (46.8,39.87) -- cycle ;
        \draw  [color={rgb, 255:red, 42; green, 191; blue, 76 }  ,draw opacity=1 ][fill={rgb, 255:red, 42; green, 191; blue, 76 }  ,fill opacity=1 ][line width=1.5]  (130.71,39.31) .. controls (130.71,37.85) and (131.93,36.67) .. (133.43,36.67) .. controls (134.94,36.67) and (136.16,37.85) .. (136.16,39.31) .. controls (136.16,40.77) and (134.94,41.95) .. (133.43,41.95) .. controls (131.93,41.95) and (130.71,40.77) .. (130.71,39.31) -- cycle ;
        \draw  [color={rgb, 255:red, 42; green, 191; blue, 76 }  ,draw opacity=1 ][dash pattern={on 1.69pt off 2.76pt}][line width=1.5]  (125.19,39.31) .. controls (125.19,34.9) and (128.88,31.32) .. (133.43,31.32) .. controls (137.99,31.32) and (141.68,34.9) .. (141.68,39.31) .. controls (141.68,43.73) and (137.99,47.31) .. (133.43,47.31) .. controls (128.88,47.31) and (125.19,43.73) .. (125.19,39.31) -- cycle ;
        \draw  [color={rgb, 255:red, 42; green, 191; blue, 76 }  ,draw opacity=1 ][dash pattern={on 1.69pt off 2.76pt}][line width=1.5]  (117.36,39.31) .. controls (117.36,30.71) and (124.56,23.74) .. (133.43,23.74) .. controls (142.31,23.74) and (149.5,30.71) .. (149.5,39.31) .. controls (149.5,47.91) and (142.31,54.89) .. (133.43,54.89) .. controls (124.56,54.89) and (117.36,47.91) .. (117.36,39.31) -- cycle ;
        \draw  [color={rgb, 255:red, 42; green, 191; blue, 76 }  ,draw opacity=1 ][fill={rgb, 255:red, 42; green, 191; blue, 76 }  ,fill opacity=1 ][line width=1.5]  (750.22,97.45) .. controls (750.22,95.05) and (752.22,93.1) .. (754.69,93.1) .. controls (757.16,93.1) and (759.16,95.05) .. (759.16,97.45) .. controls (759.16,99.86) and (757.16,101.8) .. (754.69,101.8) .. controls (752.22,101.8) and (750.22,99.86) .. (750.22,97.45) -- cycle ;
        \draw  [color={rgb, 255:red, 42; green, 191; blue, 76 }  ,draw opacity=1 ][dash pattern={on 1.69pt off 2.76pt}][line width=1.5]  (737.25,97.69) .. controls (737.25,88.18) and (745.17,80.47) .. (754.94,80.47) .. controls (764.71,80.47) and (772.63,88.18) .. (772.63,97.69) .. controls (772.63,107.21) and (764.71,114.92) .. (754.94,114.92) .. controls (745.17,114.92) and (737.25,107.21) .. (737.25,97.69) -- cycle ;
        \draw  [color={rgb, 255:red, 42; green, 191; blue, 76 }  ,draw opacity=1 ][dash pattern={on 1.69pt off 2.76pt}][line width=1.5]  (697.39,97.69) .. controls (697.39,79.02) and (715.24,63.88) .. (737.25,63.88) .. controls (759.26,63.88) and (777.1,79.02) .. (777.1,97.69) .. controls (777.1,116.37) and (759.26,131.51) .. (737.25,131.51) .. controls (715.24,131.51) and (697.39,116.37) .. (697.39,97.69) -- cycle ;
        \draw  [color={rgb, 255:red, 74; green, 144; blue, 226 }  ,draw opacity=1 ][fill={rgb, 255:red, 74; green, 144; blue, 226 }  ,fill opacity=1 ][line width=1.5]  (685.57,162.49) .. controls (685.57,159.95) and (687.68,157.9) .. (690.29,157.9) .. controls (692.89,157.9) and (695.01,159.95) .. (695.01,162.49) .. controls (695.01,165.03) and (692.89,167.09) .. (690.29,167.09) .. controls (687.68,167.09) and (685.57,165.03) .. (685.57,162.49) -- cycle ;
        \draw  [color={rgb, 255:red, 74; green, 144; blue, 226 }  ,draw opacity=1 ][dash pattern={on 1.69pt off 2.76pt}][line width=1.5]  (676.01,162.49) .. controls (676.01,154.81) and (682.4,148.59) .. (690.29,148.59) .. controls (698.17,148.59) and (704.56,154.81) .. (704.56,162.49) .. controls (704.56,170.17) and (698.17,176.39) .. (690.29,176.39) .. controls (682.4,176.39) and (676.01,170.17) .. (676.01,162.49) -- cycle ;
        \draw  [color={rgb, 255:red, 74; green, 144; blue, 226 }  ,draw opacity=1 ][dash pattern={on 1.69pt off 2.76pt}][line width=1.5]  (662.47,162.49) .. controls (662.47,147.53) and (674.92,135.4) .. (690.29,135.4) .. controls (705.65,135.4) and (718.11,147.53) .. (718.11,162.49) .. controls (718.11,177.45) and (705.65,189.58) .. (690.29,189.58) .. controls (674.92,189.58) and (662.47,177.45) .. (662.47,162.49) -- cycle ;
        
        \draw  [color={rgb, 255:red, 208; green, 2; blue, 27 }  ,draw opacity=1 ][fill={rgb, 255:red, 208; green, 2; blue, 27 }  ,fill opacity=1 ][line width=1.5]  (658.08,187.31) .. controls (658.08,183.45) and (661.2,180.33) .. (665.05,180.33) .. controls (668.9,180.33) and (672.02,183.45) .. (672.02,187.31) .. controls (672.02,191.17) and (668.9,194.3) .. (665.05,194.3) .. controls (661.2,194.3) and (658.08,191.17) .. (658.08,187.31) -- cycle ;
        \draw  [color={rgb, 255:red, 208; green, 2; blue, 27 }  ,draw opacity=1 ][dash pattern={on 1.69pt off 2.76pt}][line width=1.5]  (650.18,187.31) .. controls (650.18,179.09) and (656.84,172.42) .. (665.05,172.42) .. controls (673.26,172.42) and (679.92,179.09) .. (679.92,187.31) .. controls (679.92,195.53) and (673.26,202.2) .. (665.05,202.2) .. controls (656.84,202.2) and (650.18,195.53) .. (650.18,187.31) -- cycle ;
        \draw  [color={rgb, 255:red, 208; green, 2; blue, 27 }  ,draw opacity=1 ][dash pattern={on 1.69pt off 2.76pt}][line width=1.5]  (637.27,187.31) .. controls (637.27,171.94) and (649.71,159.49) .. (665.05,159.49) .. controls (680.39,159.49) and (692.83,171.94) .. (692.83,187.31) .. controls (692.83,202.68) and (680.39,215.14) .. (665.05,215.14) .. controls (649.71,215.14) and (637.27,202.68) .. (637.27,187.31) -- cycle ;
        \draw  [color={rgb, 255:red, 208; green, 2; blue, 27 }  ,draw opacity=1 ][dash pattern={on 1.69pt off 2.76pt}][line width=1.5]  (620.51,187.31) .. controls (620.51,162.68) and (640.45,142.7) .. (665.05,142.7) .. controls (689.65,142.7) and (709.59,162.68) .. (709.59,187.31) .. controls (709.59,211.95) and (689.65,231.92) .. (665.05,231.92) .. controls (640.45,231.92) and (620.51,211.95) .. (620.51,187.31) -- cycle ;
        \draw  [color={rgb, 255:red, 208; green, 2; blue, 27 }  ,draw opacity=1 ][dash pattern={on 1.69pt off 2.76pt}][line width=1.5]  (589.67,187.31) .. controls (589.67,145.62) and (623.42,111.82) .. (665.05,111.82) .. controls (706.68,111.82) and (740.42,145.62) .. (740.42,187.31) .. controls (740.42,229) and (706.68,262.8) .. (665.05,262.8) .. controls (623.42,262.8) and (589.67,229) .. (589.67,187.31) -- cycle ;
        \draw  [color={rgb, 255:red, 208; green, 2; blue, 27 }  ,draw opacity=1 ][dash pattern={on 1.69pt off 2.76pt}][line width=1.5]  (551.5,187.31) .. controls (551.5,124.51) and (602.34,73.59) .. (665.05,73.59) .. controls (727.76,73.59) and (778.6,124.51) .. (778.6,187.31) .. controls (778.6,250.12) and (727.76,301.03) .. (665.05,301.03) .. controls (602.34,301.03) and (551.5,250.12) .. (551.5,187.31) -- cycle ;
        
        \draw  [color={rgb, 255:red, 189; green, 16; blue, 224 }  ,draw opacity=1 ][fill={rgb, 255:red, 189; green, 16; blue, 224 }  ,fill opacity=1 ][line width=1.5]  (833.46,157.41) .. controls (833.46,152.34) and (837.68,148.23) .. (842.89,148.23) .. controls (848.1,148.23) and (852.33,152.34) .. (852.33,157.41) .. controls (852.33,162.49) and (848.1,166.6) .. (842.89,166.6) .. controls (837.68,166.6) and (833.46,162.49) .. (833.46,157.41) -- cycle ;
        \draw  [color={rgb, 255:red, 189; green, 16; blue, 224 }  ,draw opacity=1 ][dash pattern={on 1.69pt off 2.76pt}][line width=1.5]  (822.78,157.41) .. controls (822.78,146.6) and (831.79,137.83) .. (842.89,137.83) .. controls (854,137.83) and (863,146.6) .. (863,157.41) .. controls (863,168.23) and (854,177) .. (842.89,177) .. controls (831.79,177) and (822.78,168.23) .. (822.78,157.41) -- cycle ;
        \draw  [color={rgb, 255:red, 189; green, 16; blue, 224 }  ,draw opacity=1 ][dash pattern={on 1.69pt off 2.76pt}][line width=1.5]  (805.31,157.41) .. controls (805.31,137.2) and (822.14,120.81) .. (842.89,120.81) .. controls (863.65,120.81) and (880.48,137.2) .. (880.48,157.41) .. controls (880.48,177.63) and (863.65,194.01) .. (842.89,194.01) .. controls (822.14,194.01) and (805.31,177.63) .. (805.31,157.41) -- cycle ;
        \draw  [color={rgb, 255:red, 189; green, 16; blue, 224 }  ,draw opacity=1 ][dash pattern={on 1.69pt off 2.76pt}][line width=1.5]  (782.64,157.41) .. controls (782.64,125.01) and (809.61,98.74) .. (842.89,98.74) .. controls (876.17,98.74) and (903.15,125.01) .. (903.15,157.41) .. controls (903.15,189.82) and (876.17,216.09) .. (842.89,216.09) .. controls (809.61,216.09) and (782.64,189.82) .. (782.64,157.41) -- cycle ;
        
        \draw  [fill={rgb, 255:red, 155; green, 155; blue, 155 }  ,fill opacity=0.1 ][line width=1.5]  (662.47,162.49) .. controls (662.47,126.23) and (692.4,96.83) .. (729.31,96.83) .. controls (766.23,96.83) and (796.16,126.23) .. (796.16,162.49) .. controls (796.16,198.76) and (766.23,228.16) .. (729.31,228.16) .. controls (692.4,228.16) and (662.47,198.76) .. (662.47,162.49) -- cycle ;
        \draw  [color={rgb, 255:red, 42; green, 191; blue, 76 }  ,draw opacity=1 ][dash pattern={on 1.69pt off 2.76pt}][line width=1.5]  (584.49,97.71) .. controls (584.49,67.23) and (628.72,42.52) .. (683.29,42.52) .. controls (737.86,42.52) and (782.1,67.23) .. (782.1,97.71) .. controls (782.1,128.19) and (737.86,152.9) .. (683.29,152.9) .. controls (628.72,152.9) and (584.49,128.19) .. (584.49,97.71) -- cycle ;
        \draw  [color={rgb, 255:red, 42; green, 191; blue, 76 }  ,draw opacity=1 ][dash pattern={on 1.69pt off 2.76pt}][line width=1.5]  (651.49,97.82) .. controls (651.49,73.68) and (680.05,54.11) .. (715.29,54.11) .. controls (750.53,54.11) and (779.1,73.68) .. (779.1,97.82) .. controls (779.1,121.95) and (750.53,141.52) .. (715.29,141.52) .. controls (680.05,141.52) and (651.49,121.95) .. (651.49,97.82) -- cycle ;    
        
        \draw (369,39) node [scale=0.8] [align=left] {Considered subspace of the environment };
        \draw (180.71,39.31) node  [align=left] {{\footnotesize Devices}};
        \draw (137,356) node  [align=left] {\textbf{fig. a} : The effects produced by the\\device are able of modifying certain \\properties of the considered \\subspace of the environment};
        \draw (735.42,355.31) node  [align=left] {\textbf{fig. c }: Different devices allow to modify \\a given physical property of the considered \\subspace of the environment. It remains \\to select the most relevant one..};
        \draw (411,347) node  [align=left] {\textbf{fig. b} : The nomadic nature of the \\devices does not guarantee their \\availability in time and space};

\end{tikzpicture}

\caption{\label{contextual}\small Contextual factors. }
\end{figure*}
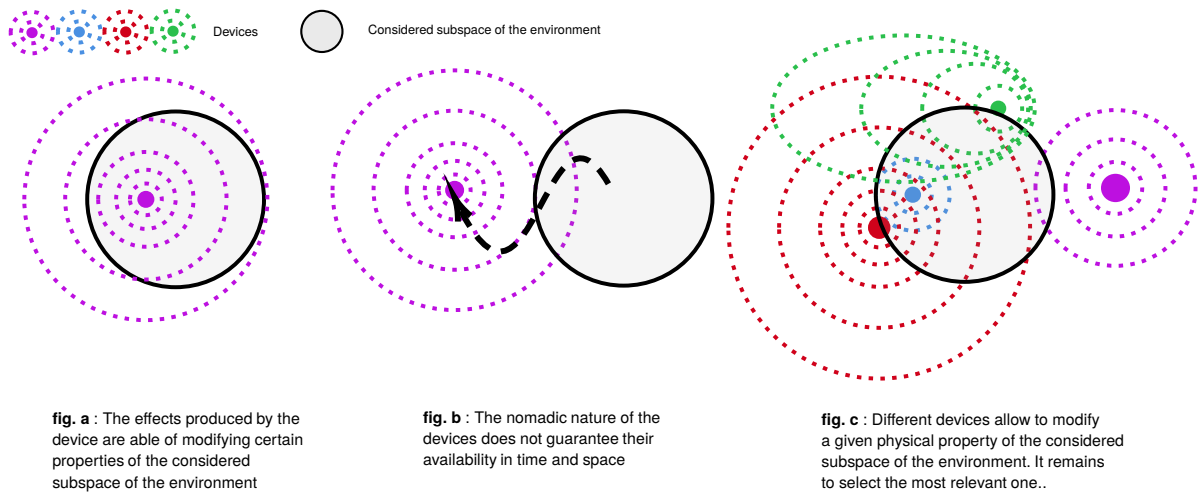

%% file: systemic.tex
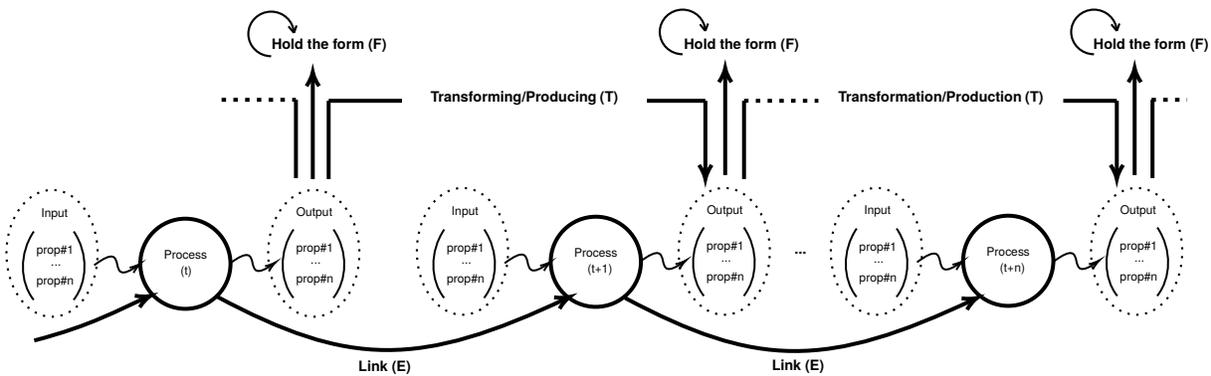
\begin{figure*}
    \centering
    \tikzset{every picture/.style={line width=0.75pt}} 
    \begin{tikzpicture}[thick,scale=0.5, every node/.style={scale=0.5}, x=0.75pt,y=0.75pt,yscale=-1,xscale=1,font=\sffamily]

    \draw  [line width=1.5]  (715,1018.5) .. controls (715,993.37) and (735.37,973) .. (760.5,973) .. controls (785.63,973) and (806,993.37) .. (806,1018.5) .. controls (806,1043.63) and (785.63,1064) .. (760.5,1064) .. controls (735.37,1064) and (715,1043.63) .. (715,1018.5) -- cycle ;
    \draw [line width=0.75]    (668.5,1020) .. controls (708.1,990.3) and (675.66,1047.34) .. (713.82,1019.37) ;
    \draw [shift={(715,1018.5)}, rotate = 503.13] [color={rgb, 255:red, 0; green, 0; blue, 0 }  ][line width=0.75]    (10.93,-3.29) .. controls (6.95,-1.4) and (3.31,-0.3) .. (0,0) .. controls (3.31,0.3) and (6.95,1.4) .. (10.93,3.29)   ;
    
    \draw [line width=0.75]    (607.5,1055) .. controls (593.37,1037.38) and (592.62,1007.88) .. (607.62,985.88) ;

    \draw [line width=0.75]    (647.75,1055) .. controls (661.62,1037.38) and (663.12,1008.38) .. (647.87,985.88) ;

    \draw [color={rgb, 255:red, 0; green, 0; blue, 0 }  ,draw opacity=1 ][line width=0.75]    (805,1018.5) .. controls (844.6,988.8) and (812.16,1045.84) .. (850.32,1017.87) ;
    \draw [shift={(851.5,1017)}, rotate = 503.13] [color={rgb, 255:red, 0; green, 0; blue, 0 }  ,draw opacity=1 ][line width=0.75]    (10.93,-3.29) .. controls (6.95,-1.4) and (3.31,-0.3) .. (0,0) .. controls (3.31,0.3) and (6.95,1.4) .. (10.93,3.29)   ;
    
    \draw [color={rgb, 255:red, 0; green, 0; blue, 0 }  ,draw opacity=1 ][line width=0.75]    (869.5,1053) .. controls (855.37,1035.38) and (854.62,1005.88) .. (869.62,983.88) ;

    \draw [color={rgb, 255:red, 0; green, 0; blue, 0 }  ,draw opacity=1 ][line width=0.75]    (909.75,1053) .. controls (923.62,1035.38) and (925.12,1006.38) .. (909.87,983.88) ;

    \draw  [dash pattern={on 0.84pt off 2.51pt}] (580,1008) .. controls (580,972.65) and (600.59,944) .. (626,944) .. controls (651.41,944) and (672,972.65) .. (672,1008) .. controls (672,1043.35) and (651.41,1072) .. (626,1072) .. controls (600.59,1072) and (580,1043.35) .. (580,1008) -- cycle ;
    \draw  [dash pattern={on 0.84pt off 2.51pt}] (844,1008) .. controls (844,972.65) and (864.59,944) .. (890,944) .. controls (915.41,944) and (936,972.65) .. (936,1008) .. controls (936,1043.35) and (915.41,1072) .. (890,1072) .. controls (864.59,1072) and (844,1043.35) .. (844,1008) -- cycle ;
    \draw  [line width=1.5]  (1545,1015.5) .. controls (1545,990.37) and (1565.37,970) .. (1590.5,970) .. controls (1615.63,970) and (1636,990.37) .. (1636,1015.5) .. controls (1636,1040.63) and (1615.63,1061) .. (1590.5,1061) .. controls (1565.37,1061) and (1545,1040.63) .. (1545,1015.5) -- cycle ;
    \draw [line width=0.75]    (1498.5,1017) .. controls (1538.1,987.3) and (1505.66,1044.34) .. (1543.82,1016.37) ;
    \draw [shift={(1545,1015.5)}, rotate = 503.13] [color={rgb, 255:red, 0; green, 0; blue, 0 }  ][line width=0.75]    (10.93,-3.29) .. controls (6.95,-1.4) and (3.31,-0.3) .. (0,0) .. controls (3.31,0.3) and (6.95,1.4) .. (10.93,3.29)   ;
    
    \draw [line width=0.75]    (1437.5,1052) .. controls (1423.37,1034.38) and (1422.62,1004.88) .. (1437.62,982.88) ;

    \draw [line width=0.75]    (1477.75,1052) .. controls (1491.62,1034.38) and (1493.12,1005.38) .. (1477.87,982.88) ;

    \draw [color={rgb, 255:red, 0; green, 0; blue, 0 }  ,draw opacity=1 ][line width=0.75]    (1636,1016.5) .. controls (1675.6,986.8) and (1643.16,1043.84) .. (1681.32,1015.87) ;
    \draw [shift={(1682.5,1015)}, rotate = 503.13] [color={rgb, 255:red, 0; green, 0; blue, 0 }  ,draw opacity=1 ][line width=0.75]    (10.93,-3.29) .. controls (6.95,-1.4) and (3.31,-0.3) .. (0,0) .. controls (3.31,0.3) and (6.95,1.4) .. (10.93,3.29)   ;
    
    \draw [color={rgb, 255:red, 0; green, 0; blue, 0 }  ,draw opacity=1 ][line width=0.75]    (1700.5,1051) .. controls (1686.37,1033.38) and (1685.62,1003.88) .. (1700.62,981.88) ;

    \draw [color={rgb, 255:red, 0; green, 0; blue, 0 }  ,draw opacity=1 ][line width=0.75]    (1740.75,1051) .. controls (1754.62,1033.38) and (1756.12,1004.38) .. (1740.87,981.88) ;

    \draw  [dash pattern={on 0.84pt off 2.51pt}] (1411.5,1006) .. controls (1411.5,970.65) and (1432.09,942) .. (1457.5,942) .. controls (1482.91,942) and (1503.5,970.65) .. (1503.5,1006) .. controls (1503.5,1041.35) and (1482.91,1070) .. (1457.5,1070) .. controls (1432.09,1070) and (1411.5,1041.35) .. (1411.5,1006) -- cycle ;
    \draw [line width=1.5]    (1285,854) -- (1285,930) ;
    \draw [shift={(1285,933)}, rotate = 270] [color={rgb, 255:red, 0; green, 0; blue, 0 }  ][line width=1.5]    (14.21,-6.37) .. controls (9.04,-2.99) and (4.3,-0.87) .. (0,0) .. controls (4.3,0.87) and (9.04,2.99) .. (14.21,6.37)   ;
    
    \draw [line width=1.5]    (905,854) -- (979,854) ;

    \draw [line width=1.5]    (1226,854) -- (1285,854) ;

    \draw [line width=1.5]    (905,854) -- (905,933) ;

    \draw  [dash pattern={on 0.84pt off 2.51pt}] (1674,1006) .. controls (1674,970.65) and (1694.59,942) .. (1720,942) .. controls (1745.41,942) and (1766,970.65) .. (1766,1006) .. controls (1766,1041.35) and (1745.41,1070) .. (1720,1070) .. controls (1694.59,1070) and (1674,1041.35) .. (1674,1006) -- cycle ;
    \draw  [line width=1.5]  (1129,1016.5) .. controls (1129,991.37) and (1149.37,971) .. (1174.5,971) .. controls (1199.63,971) and (1220,991.37) .. (1220,1016.5) .. controls (1220,1041.63) and (1199.63,1062) .. (1174.5,1062) .. controls (1149.37,1062) and (1129,1041.63) .. (1129,1016.5) -- cycle ;
    \draw [line width=0.75]    (1082.5,1018) .. controls (1122.1,988.3) and (1089.66,1045.34) .. (1127.82,1017.37) ;
    \draw [shift={(1129,1016.5)}, rotate = 503.13] [color={rgb, 255:red, 0; green, 0; blue, 0 }  ][line width=0.75]    (10.93,-3.29) .. controls (6.95,-1.4) and (3.31,-0.3) .. (0,0) .. controls (3.31,0.3) and (6.95,1.4) .. (10.93,3.29)   ;
    
    \draw [line width=0.75]    (1021.5,1053) .. controls (1007.37,1035.38) and (1006.62,1005.88) .. (1021.62,983.88) ;

    \draw [line width=0.75]    (1061.75,1053) .. controls (1075.62,1035.38) and (1077.12,1006.38) .. (1061.87,983.88) ;

    \draw [color={rgb, 255:red, 0; green, 0; blue, 0 }  ,draw opacity=1 ][line width=0.75]    (1219,1016.5) .. controls (1258.6,986.8) and (1226.16,1043.84) .. (1264.32,1015.87) ;
    \draw [shift={(1265.5,1015)}, rotate = 503.13] [color={rgb, 255:red, 0; green, 0; blue, 0 }  ,draw opacity=1 ][line width=0.75]    (10.93,-3.29) .. controls (6.95,-1.4) and (3.31,-0.3) .. (0,0) .. controls (3.31,0.3) and (6.95,1.4) .. (10.93,3.29)   ;
    
    \draw [color={rgb, 255:red, 0; green, 0; blue, 0 }  ,draw opacity=1 ][line width=0.75]    (1283.5,1051) .. controls (1269.37,1033.38) and (1268.62,1003.88) .. (1283.62,981.88) ;

    \draw [color={rgb, 255:red, 0; green, 0; blue, 0 }  ,draw opacity=1 ][line width=0.75]    (1323.75,1051) .. controls (1337.62,1033.38) and (1339.12,1004.38) .. (1323.87,981.88) ;

    \draw  [dash pattern={on 0.84pt off 2.51pt}] (994,1006) .. controls (994,970.65) and (1014.59,942) .. (1040,942) .. controls (1065.41,942) and (1086,970.65) .. (1086,1006) .. controls (1086,1041.35) and (1065.41,1070) .. (1040,1070) .. controls (1014.59,1070) and (994,1041.35) .. (994,1006) -- cycle ;
    \draw  [dash pattern={on 0.84pt off 2.51pt}] (1258,1006) .. controls (1258,970.65) and (1278.59,942) .. (1304,942) .. controls (1329.41,942) and (1350,970.65) .. (1350,1006) .. controls (1350,1041.35) and (1329.41,1070) .. (1304,1070) .. controls (1278.59,1070) and (1258,1041.35) .. (1258,1006) -- cycle ;
    \draw [line width=1.5]    (1700,854) -- (1700,930) ;
    \draw [shift={(1700,933)}, rotate = 270] [color={rgb, 255:red, 0; green, 0; blue, 0 }  ][line width=1.5]    (14.21,-6.37) .. controls (9.04,-2.99) and (4.3,-0.87) .. (0,0) .. controls (4.3,0.87) and (9.04,2.99) .. (14.21,6.37)   ;
    
    \draw [line width=1.5]  [dash pattern={on 1.69pt off 2.76pt}]  (1324,854) -- (1399,854) ;

    \draw [line width=1.5]    (1646,854) -- (1700,854) ;

    \draw [line width=1.5]    (1324,854) -- (1324,933) ;

    \draw [line width=1.5]    (793,1052) .. controls (925.34,1122.65) and (996.29,1123.99) .. (1145.74,1053.07) ;
    \draw [shift={(1148,1052)}, rotate = 514.51] [color={rgb, 255:red, 0; green, 0; blue, 0 }  ][line width=1.5]    (14.21,-6.37) .. controls (9.04,-2.99) and (4.3,-0.87) .. (0,0) .. controls (4.3,0.87) and (9.04,2.99) .. (14.21,6.37)   ;
    
    \draw [line width=1.5]    (1204,1053) .. controls (1336.34,1123.65) and (1407.29,1124.99) .. (1556.74,1054.07) ;
    \draw [shift={(1559,1053)}, rotate = 514.51] [color={rgb, 255:red, 0; green, 0; blue, 0 }  ][line width=1.5]    (14.21,-6.37) .. controls (9.04,-2.99) and (4.3,-0.87) .. (0,0) .. controls (4.3,0.87) and (9.04,2.99) .. (14.21,6.37)   ;
    
    \draw [line width=1.5]    (889,828) -- (889,933) ;
    
    \draw [shift={(889,825)}, rotate = 90] [color={rgb, 255:red, 0; green, 0; blue, 0 }  ][line width=1.5]    (14.21,-6.37) .. controls (9.04,-2.99) and (4.3,-0.87) .. (0,0) .. controls (4.3,0.87) and (9.04,2.99) .. (14.21,6.37)   ;
    \draw [line width=1.5]    (1305,824) -- (1305,929) ;
    
    \draw [shift={(1305,821)}, rotate = 90] [color={rgb, 255:red, 0; green, 0; blue, 0 }  ][line width=1.5]    (14.21,-6.37) .. controls (9.04,-2.99) and (4.3,-0.87) .. (0,0) .. controls (4.3,0.87) and (9.04,2.99) .. (14.21,6.37)   ;
    \draw [line width=1.5]    (1718,824) -- (1718,929) ;
    
    \draw [shift={(1718,821)}, rotate = 90] [color={rgb, 255:red, 0; green, 0; blue, 0 }  ][line width=1.5]    (14.21,-6.37) .. controls (9.04,-2.99) and (4.3,-0.87) .. (0,0) .. controls (4.3,0.87) and (9.04,2.99) .. (14.21,6.37)   ;
    \draw [line width=1.5]    (608,1096) .. controls (667.48,1080.4) and (690.83,1065.75) .. (722.54,1052.05) ;
    \draw [shift={(725,1051)}, rotate = 517.01] [color={rgb, 255:red, 0; green, 0; blue, 0 }  ][line width=1.5]    (14.21,-6.37) .. controls (9.04,-2.99) and (4.3,-0.87) .. (0,0) .. controls (4.3,0.87) and (9.04,2.99) .. (14.21,6.37)   ;
    
    \draw [line width=1.5]  [dash pattern={on 1.69pt off 2.76pt}]  (1735,853) -- (1771,853) ;

    \draw [line width=1.5]    (1736,853) -- (1736,932) ;

    \draw [line width=1.5]  [dash pattern={on 1.69pt off 2.76pt}]  (797,853) -- (872,853) ;

    \draw [line width=1.5]    (872,853) -- (872,932) ;

    \draw  [draw opacity=0] (847,809) .. controls (847,809) and (847,809) .. (847,809) .. controls (834.3,809) and (824,798.93) .. (824,786.5) .. controls (824,774.07) and (834.3,764) .. (847,764) .. controls (859.34,764) and (869.41,773.51) .. (869.98,785.44) -- (847,786.5) -- cycle ; \draw   (847,809) .. controls (847,809) and (847,809) .. (847,809) .. controls (834.3,809) and (824,798.93) .. (824,786.5) .. controls (824,774.07) and (834.3,764) .. (847,764) .. controls (859.34,764) and (869.41,773.51) .. (869.98,785.44) ;
    \draw    (864.5,778.8) -- (869.98,785.44) ;

    \draw    (869.98,785.44) -- (872.7,778) ;

    \draw  [draw opacity=0] (1262,807) .. controls (1262,807) and (1262,807) .. (1262,807) .. controls (1249.3,807) and (1239,796.93) .. (1239,784.5) .. controls (1239,772.07) and (1249.3,762) .. (1262,762) .. controls (1274.34,762) and (1284.41,771.51) .. (1284.98,783.44) -- (1262,784.5) -- cycle ; \draw   (1262,807) .. controls (1262,807) and (1262,807) .. (1262,807) .. controls (1249.3,807) and (1239,796.93) .. (1239,784.5) .. controls (1239,772.07) and (1249.3,762) .. (1262,762) .. controls (1274.34,762) and (1284.41,771.51) .. (1284.98,783.44) ;
    \draw    (1279.5,776.8) -- (1284.98,783.44) ;

    \draw    (1284.98,783.44) -- (1287.7,776) ;

    \draw  [draw opacity=0] (1677,807) .. controls (1677,807) and (1677,807) .. (1677,807) .. controls (1664.3,807) and (1654,796.93) .. (1654,784.5) .. controls (1654,772.07) and (1664.3,762) .. (1677,762) .. controls (1689.34,762) and (1699.41,771.51) .. (1699.98,783.44) -- (1677,784.5) -- cycle ; \draw   (1677,807) .. controls (1677,807) and (1677,807) .. (1677,807) .. controls (1664.3,807) and (1654,796.93) .. (1654,784.5) .. controls (1654,772.07) and (1664.3,762) .. (1677,762) .. controls (1689.34,762) and (1699.41,771.51) .. (1699.98,783.44) ;
    \draw    (1694.5,776.8) -- (1699.98,783.44) ;

    \draw    (1699.98,783.44) -- (1702.7,776) ;

    \draw (760.5,1018.5) node  [align=left] {Process\\ \ \ \ \ \ (t)};
    \draw (628.5,968.5) node  [align=left] {Input};
    \draw (1590.5,1015.5) node  [align=left] {Process\\ \ \ \ \ (t+n)};
    \draw (1458.5,965.5) node  [align=left] {Input};
    \draw (890.5,967.5) node [color={rgb, 255:red, 0; green, 0; blue, 0 }  ,opacity=1 ] [align=left] {Output};
    \draw (891,1003) node [color={rgb, 255:red, 0; green, 0; blue, 0 }  ,opacity=1 ] [align=left] {prop\#1};
    \draw (892,1018) node [color={rgb, 255:red, 0; green, 0; blue, 0 }  ,opacity=1 ] [align=left] {...};
    \draw (891,1035) node [color={rgb, 255:red, 0; green, 0; blue, 0 }  ,opacity=1 ] [align=left] {prop\#n};
    \draw (1721.5,965.5) node [color={rgb, 255:red, 0; green, 0; blue, 0 }  ,opacity=1 ] [align=left] {Output};
    \draw (1722,1001) node [color={rgb, 255:red, 0; green, 0; blue, 0 }  ,opacity=1 ] [align=left] {prop\#1};
    \draw (1723,1016) node [color={rgb, 255:red, 0; green, 0; blue, 0 }  ,opacity=1 ] [align=left] {...};
    \draw (1722,1033) node [color={rgb, 255:red, 0; green, 0; blue, 0 }  ,opacity=1 ] [align=left] {prop\#n};
    \draw (1102.5,851.5) node  [align=left] {{\large \textbf{Transforming/Producing (T)}}};
    \draw (1174.5,1016.5) node  [align=left] {Process\\ \ \ \ \ (t+1)};
    \draw (1042.5,966.5) node  [align=left] {Input};
    \draw (1304.5,965.5) node [color={rgb, 255:red, 0; green, 0; blue, 0 }  ,opacity=1 ] [align=left] {Output};
    \draw (1305,1001) node [color={rgb, 255:red, 0; green, 0; blue, 0 }  ,opacity=1 ] [align=left] {prop\#1};
    \draw (1306,1016) node [color={rgb, 255:red, 0; green, 0; blue, 0 }  ,opacity=1 ] [align=left] {...};
    \draw (1305,1033) node [color={rgb, 255:red, 0; green, 0; blue, 0 }  ,opacity=1 ] [align=left] {prop\#n};
    \draw (1381,1007) node [scale=1.2] [align=left] {\textbf{...}};
    \draw (1522.5,851.5) node  [align=left] {{\large \textbf{Transformation/Production (T)}}};
    \draw (961.5,1123.5) node  [align=left] {\textbf{{\large Link (E)}}};
    \draw (1378.5,1122.5) node  [align=left] {\textbf{{\large Link (E)}}};
    \draw (905.5,797.5) node  [align=left] {{\large \textbf{Hold the form (F)}}};
    \draw (1321.5,797.5) node  [align=left] {{\large \textbf{Hold the form (F)}}};
    \draw (1734.5,797.5) node  [align=left] {{\large \textbf{Hold the form (F)}}};
    \draw (1043,1003) node  [align=left] {prop\#1};
    \draw (1044,1018) node  [align=left] {...};
    \draw (1043,1035) node  [align=left] {prop\#n};
    \draw (1459,1002) node  [align=left] {prop\#1};
    \draw (1460,1017) node  [align=left] {...};
    \draw (1459,1034) node  [align=left] {prop\#n};
    \draw (629,1005) node  [align=left] {prop\#1};
    \draw (630,1020) node  [align=left] {...};
    \draw (629,1037) node  [align=left] {prop\#n};
    
    \end{tikzpicture}
    \caption{\label{systemic} Systemic modelling consists in considering a complex system as the conjunction of three actions: that of transforming/producing (T) over time, that of linking (E) in space and that of maintaining an identifiable form (F) characterized by a vector whose components correspond to certain values of physical properties.}
    \end{figure*}
    
    

%% file: gibert.tex
 
\tikzset{
pattern size/.store in=\mcSize, 
pattern size = 5pt,
pattern thickness/.store in=\mcThickness, 
pattern thickness = 0.3pt,
pattern radius/.store in=\mcRadius, 
pattern radius = 1pt}
\makeatletter
\pgfutil@ifundefined{pgf@pattern@name@_vbtt3jhbq}{
\pgfdeclarepatternformonly[\mcThickness,\mcSize]{_vbtt3jhbq}
{\pgfqpoint{0pt}{0pt}}
{\pgfpoint{\mcSize+\mcThickness}{\mcSize+\mcThickness}}
{\pgfpoint{\mcSize}{\mcSize}}
{
\pgfsetcolor{\tikz@pattern@color}
\pgfsetlinewidth{\mcThickness}
\pgfpathmoveto{\pgfqpoint{0pt}{0pt}}
\pgfpathlineto{\pgfpoint{\mcSize+\mcThickness}{\mcSize+\mcThickness}}
\pgfusepath{stroke}
}}
\makeatother

 
\tikzset{
pattern size/.store in=\mcSize, 
pattern size = 5pt,
pattern thickness/.store in=\mcThickness, 
pattern thickness = 0.3pt,
pattern radius/.store in=\mcRadius, 
pattern radius = 1pt}
\makeatletter
\pgfutil@ifundefined{pgf@pattern@name@_e4drsqpfu}{
\pgfdeclarepatternformonly[\mcThickness,\mcSize]{_e4drsqpfu}
{\pgfqpoint{0pt}{0pt}}
{\pgfpoint{\mcSize+\mcThickness}{\mcSize+\mcThickness}}
{\pgfpoint{\mcSize}{\mcSize}}
{
\pgfsetcolor{\tikz@pattern@color}
\pgfsetlinewidth{\mcThickness}
\pgfpathmoveto{\pgfqpoint{0pt}{0pt}}
\pgfpathlineto{\pgfpoint{\mcSize+\mcThickness}{\mcSize+\mcThickness}}
\pgfusepath{stroke}
}}
\makeatother

 
\tikzset{
pattern size/.store in=\mcSize, 
pattern size = 5pt,
pattern thickness/.store in=\mcThickness, 
pattern thickness = 0.3pt,
pattern radius/.store in=\mcRadius, 
pattern radius = 1pt}
\makeatletter
\pgfutil@ifundefined{pgf@pattern@name@_3u5q116iy}{
\pgfdeclarepatternformonly[\mcThickness,\mcSize]{_3u5q116iy}
{\pgfqpoint{0pt}{0pt}}
{\pgfpoint{\mcSize+\mcThickness}{\mcSize+\mcThickness}}
{\pgfpoint{\mcSize}{\mcSize}}
{
\pgfsetcolor{\tikz@pattern@color}
\pgfsetlinewidth{\mcThickness}
\pgfpathmoveto{\pgfqpoint{0pt}{0pt}}
\pgfpathlineto{\pgfpoint{\mcSize+\mcThickness}{\mcSize+\mcThickness}}
\pgfusepath{stroke}
}}
\makeatother

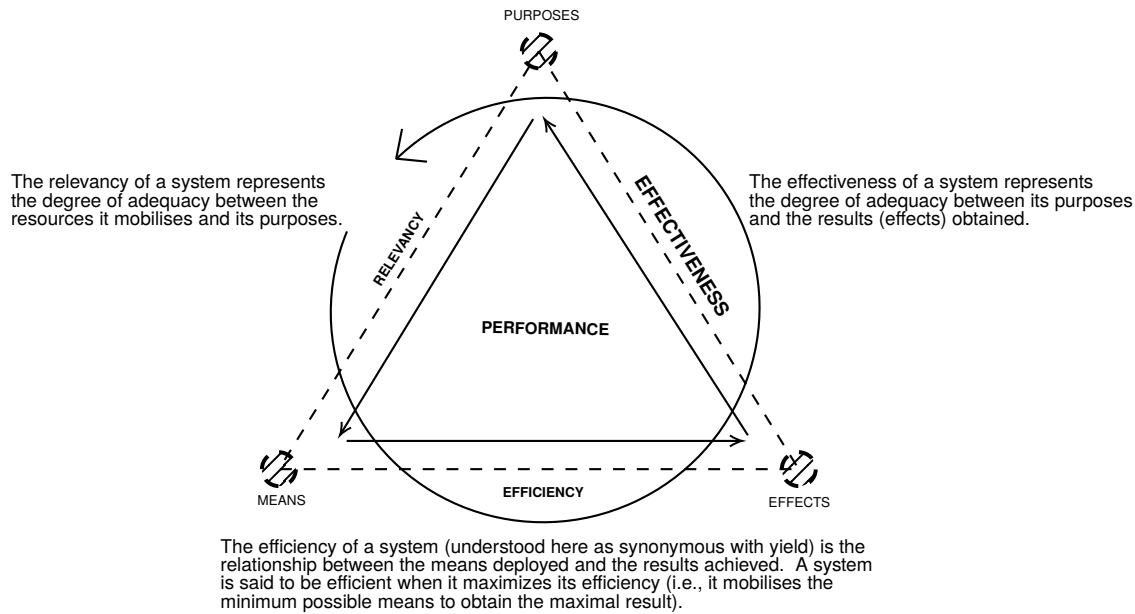
\begin{figure*}
    \centering
    
    \tikzset{every picture/.style={line width=0.75pt}} 
    \begin{tikzpicture}[thick,scale=0.57, every node/.style={scale=0.57}, x=0.75pt,y=0.75pt,yscale=-1,xscale=1,font=\sffamily]

\draw  [dash pattern={on 4.5pt off 4.5pt}] (701.5,750.81) -- (931.5,1120.81) -- (471.5,1120.81) -- cycle ;
\draw    (696.5,810.81) -- (526.54,1089.1) ;
\draw [shift={(525.5,1090.81)}, rotate = 301.40999999999997] [color={rgb, 255:red, 0; green, 0; blue, 0 }  ][line width=0.75]    (10.93,-4.9) .. controls (6.95,-2.3) and (3.31,-0.67) .. (0,0) .. controls (3.31,0.67) and (6.95,2.3) .. (10.93,4.9)   ;

\draw    (531.5,1095.81) -- (879.5,1095.81) ;
\draw [shift={(881.5,1095.81)}, rotate = 180] [color={rgb, 255:red, 0; green, 0; blue, 0 }  ][line width=0.75]    (10.93,-4.9) .. controls (6.95,-2.3) and (3.31,-0.67) .. (0,0) .. controls (3.31,0.67) and (6.95,2.3) .. (10.93,4.9)   ;

\draw    (886.5,1090.81) -- (707.58,812.49) ;
\draw [shift={(706.5,810.81)}, rotate = 417.26] [color={rgb, 255:red, 0; green, 0; blue, 0 }  ][line width=0.75]    (10.93,-4.9) .. controls (6.95,-2.3) and (3.31,-0.67) .. (0,0) .. controls (3.31,0.67) and (6.95,2.3) .. (10.93,4.9)   ;

\draw  [pattern=_vbtt3jhbq,pattern size=6pt,pattern thickness=0.75pt,pattern radius=0pt, pattern color={rgb, 255:red, 0; green, 0; blue, 0}][dash pattern={on 5.63pt off 4.5pt}][line width=1.5]  (686.5,750.81) .. controls (686.5,742.52) and (693.22,735.81) .. (701.5,735.81) .. controls (709.78,735.81) and (716.5,742.52) .. (716.5,750.81) .. controls (716.5,759.09) and (709.78,765.81) .. (701.5,765.81) .. controls (693.22,765.81) and (686.5,759.09) .. (686.5,750.81) -- cycle ;
\draw  [pattern=_e4drsqpfu,pattern size=6pt,pattern thickness=0.75pt,pattern radius=0pt, pattern color={rgb, 255:red, 0; green, 0; blue, 0}][dash pattern={on 5.63pt off 4.5pt}][line width=1.5]  (456.5,1120.81) .. controls (456.5,1112.52) and (463.22,1105.81) .. (471.5,1105.81) .. controls (479.78,1105.81) and (486.5,1112.52) .. (486.5,1120.81) .. controls (486.5,1129.09) and (479.78,1135.81) .. (471.5,1135.81) .. controls (463.22,1135.81) and (456.5,1129.09) .. (456.5,1120.81) -- cycle ;
\draw  [pattern=_3u5q116iy,pattern size=6pt,pattern thickness=0.75pt,pattern radius=0pt, pattern color={rgb, 255:red, 0; green, 0; blue, 0}][dash pattern={on 5.63pt off 4.5pt}][line width=1.5]  (916.5,1120.81) .. controls (916.5,1112.52) and (923.22,1105.81) .. (931.5,1105.81) .. controls (939.78,1105.81) and (946.5,1112.52) .. (946.5,1120.81) .. controls (946.5,1129.09) and (939.78,1135.81) .. (931.5,1135.81) .. controls (923.22,1135.81) and (916.5,1129.09) .. (916.5,1120.81) -- cycle ;
\draw  [draw opacity=0] (575.19,846.34) .. controls (590.47,831.53) and (608.51,819) .. (628.99,809.61) .. controls (724.43,765.83) and (836.77,806.6) .. (879.91,900.66) .. controls (923.06,994.72) and (880.67,1106.46) .. (785.23,1150.24) .. controls (689.79,1194.01) and (577.45,1153.25) .. (534.3,1059.18) .. controls (511.9,1010.33) and (512.56,956.71) .. (531.75,910.32) -- (707.11,979.92) -- cycle ; \draw   (575.19,846.34) .. controls (590.47,831.53) and (608.51,819) .. (628.99,809.61) .. controls (724.43,765.83) and (836.77,806.6) .. (879.91,900.66) .. controls (923.06,994.72) and (880.67,1106.46) .. (785.23,1150.24) .. controls (689.79,1194.01) and (577.45,1153.25) .. (534.3,1059.18) .. controls (511.9,1010.33) and (512.56,956.71) .. (531.75,910.32) ;
\draw    (580.5,819.5) -- (575.19,846.34) ;

\draw    (575.31,845.85) -- (602.53,848.55) ;

\draw (704,718.5) node  [align=left] {PURPOSES};
\draw (474,1148.5) node  [align=left] {MEANS};
\draw (932,1149.81) node  [align=left] {EFFECTS};
\draw (705.5,1141.5) node  [align=left] {\textbf{EFFICIENCY}};
\draw (577.73,925.72) node [rotate=-302.25] [align=left] {\textbf{RELEVANCY}};
\draw (707.11,995.92) node [rotate=-0.83] [align=left] {\textbf{{\large PERFORMANCE}}};
\draw (828.49,922.74) node [scale=1.44,rotate=-58] [align=left] {\textbf{EFFECTIVENESS}};
\draw (381.5,884) node  [align=left] {\Large The relevancy of a system represents \\\Large the degree of adequacy between the \\ \Large resources it mobilises and its purposes.};
\draw (708.5,1215) node  [align=left] {\Large The efficiency of a system (understood here as synonymous with yield) is the\\\Large relationship between the means deployed and the results achieved. \ A system \\\Large is said to be efficient when it maximizes its efficiency (i.e., it mobilises the\\ \Large minimum possible means to obtain the maximal result).};
\draw (1057.5,884) node  [align=left] {\Large The effectiveness of a system represents\\\Large the degree of adequacy between its purposes\\\Large and the results (effects) obtained.};

\end{tikzpicture}

\caption{\label{gibert}Gibert's performance model \cite{gibert1980controle}. }
\end{figure*}